\documentclass[12pt]{article}

\usepackage{amsmath,amsthm,amssymb}
\usepackage{mathtools}
\usepackage{mathrsfs}
\usepackage{graphicx}
\usepackage{float}
\usepackage{lmodern,microtype}
\usepackage{enumitem}
\usepackage{xcolor}

\usepackage[margin=1in]{geometry}

\usepackage[round]{natbib}
\newcommand{\cmmnt}[1]{}
\usepackage{setspace}
\usepackage{tabularx}
\usepackage{subcaption}
\usepackage{nicefrac}
\usepackage{etoolbox}
\usepackage{xcolor}
\usepackage{tikz}
\usepackage{hyperref}
\usepackage{biolinum}

\usepackage{blkarray}

\usepackage[pagewise]{lineno}
\newcommand{\enremark}{\hfill$\circ$}
\usepackage[most]{tcolorbox} 

\theoremstyle{remark} 

\def\co{{\rm co}\hskip 1pt}

\def\proj{{\rm proj}}
\def\proj0{{\rm proj}_{\co M_0}}

\newtheorem{theorem}{Theorem}
{\end{theorem}\vskip.2cm}

\newtheorem{defini}{Definition}
{\end{defini}\vskip.2cm}

\newtheorem{cla}{Claim}
{\end{cla}\vskip.2cm}

\newtheorem{remark}{\textbf{Remark}}
{\end{remark}\vskip.3cm}

\theoremstyle{definition}
\newtheorem{mydef}{Definition}
\newtheorem{result}{Result}
\newtheorem{proposi}{Proposition} 
\newtheorem{assumption}{Assumption}
\newtheorem{corollary}{Corollary}
\newtheorem{conjecture}{Conjecture}
\newtheorem{lemma}{Lemma}
{\end{lemma}\vskip.2cm}

\usepackage{tikz}
\usetikzlibrary{positioning}

\usepackage[english]{isodate}
\usepackage{booktabs,tabularx}
\graphicspath{ {./Image/} }

\title{When Is Degree Enough? 
\\
\large{Bounds on Degree-Eigenvector Misalignment in Assortative Structured Networks}}
\author{Sreerag Puravankara\thanks{Economics Area, Indian Institute of Management Kozhikode, Kerala 673 570, India.} \and  Vipin P. Veetil\footnotemark[2]}
          
\date{\small\printdayoff\today}

\begin{document}

\maketitle

\begin{abstract}
\setstretch{1.35}
A tight alignment between the degree vector and the leading eigenvector arises naturally in networks with neutral degree mixing and the absence of local structures. Many real-world networks, however, violate both conditions. We derive bounds on the divergence between the degree vector and the eigenvector in networks with degree assortativity and local mesoscopic structures such as communities, core-peripheries, and cycles. Our approach is constructive. We design sufficiently general degree-preserving rewiring algorithms which begins from a neutral benchmark  to monotonically change assortativity and the strength of local structures, with each step inducing a perturbation of the adjacency matrix.  Using the Stewart--Sun Perturbation Bound, together with explicit spectral-norm control of the rewiring steps, we derive upper bounds on the angle between the eigenvector and the degree vector for modest levels of assortativity and local structures.  Our analytical bounds delineate regions of `spectral safety' in which a node's degree can be used as a reliable measure of its systemic importance in real-world networks. We also substantiate our analytical bounds with numerical simulations that compute the exact angles of deviation.
\end{abstract}
\vspace{1cm}
\textbf{Keywords}: Eigenvector Localization, Steward-Sun Perturbation Bound, Assortativity, Communities, Core-Periphery, Cycles, Degree-Preserving Rewiring.
\newpage

\setstretch{1.35}
\section{Introduction}
Over the last quarter century, many disciplines---including the social and biological sciences---have embraced the idea that a node's systemic importance depends not only on how many neighbors it has, but also on where it sits in the network as a whole. In such settings, eigenvector-based measures of centrality provide a natural notion of influence or vulnerability. In practice, however, computing eigenvector centrality requires essentially complete knowledge of the network, which is rarely available. By contrast, degree information is often observed or can be estimated reasonably well. Which is why applied researchers sometimes use the degree vector as a proxy for the leading eigenvector. This is entirely reasonable when the two are close. Unfortunately, however, the conditions under which degree and eigenvector coincide are quite restrictive. The network must not only exhibit no assortativity,  it must also be devoid of all local structures, including communities, core-periphery patterns, and directed cycles. One would be hard-pressed to find a real-world network that meets these conditions. Firm buyer--seller networks, for instance, exhibit a clear core-periphery structure. Networks of disease transmission sit inside rich social graphs with intricate communities and friendship cycles\footnote{See \citet{SalatheJones2010Community} and \citet{VolzMillerGalvaniMeyers2011Clustered} for evidence on local structures within the social networks of epidemiological transmission. See \citet{Chakraborty2018} and \citet{Bacilieri2025} for empirical evidence on assortativity and local structures in firm buyer-seller networks.}. In such circumstances, the leading eigenvector can diverge from the degree profile, with the degree vector becoming a progressively worse proxy as the network departs from the neutral benchmark. Ideally, we would like to know how large this divergence is  before using the degree vector as a proxy for the eigenvector. Which is precisely the question we address in this paper.  We derive explicit bounds on the divergence between the eigenvector and the degree vector when the network departs from the neutral benchmark by modest amounts of assortativity, community structure, core-periphery pattern, and directed cycles. While our primary focus is on unweighted graphs, we also demonstrate how this framework naturally generalizes to weighted directed networks.
\\
\indent
Nearly all of what is said in this paper is built upon the Stewart-Sun Perturbation Theorem, which tells us how much the eigenvector can move under a perturbation of the adjacency matrix. In its simplest form, the theorem states that if $\mathbf A$ is diagonalizable with a simple leading eigenvalue and we form a perturbed matrix $\widetilde{\mathbf A} := \mathbf A + \boldsymbol{\Omega}$, then the sine of the angle between the corresponding leading eigenvectors is bounded above by a constant times $\|\boldsymbol{\Omega}\|_2$, the spectral norm of the perturbation. Taken at face value, this is a statement about eigenvectors before and after perturbation. It says nothing about degree vectors. Note, however, that under the neutral benchmark, the leading eigenvector of $\mathbf A$ is well approximated by the degree vector. Therefore, if we restrict perturbations to degree-preserving rewirings, the Stewart-Sun bound on the distance between the eigenvectors of $\mathbf A$ and $\widetilde{\mathbf A}$ is equivalent to a bound on the distance between the degree vector and the eigenvector of the perturbed graph $\widetilde{\mathbf A}$.  We develop this insight by constructing four families of degree-preserving rewiring schemes that progressively introduce (i) assortative mixing, (ii) community structures, (iii) core-periphery patterns, and (iv)  directed cycles. For each scheme we explicitly bound $\|\boldsymbol{\Omega}\|_2$ in terms of standard scalar measures of these features---Newman's assortativity coefficient for degree correlations, a modularity-type quantity for communities, a core-periphery contrast parameter, and normalized counts of cycles. Substituting these into the Stewart--Sun inequality yields explicit, interpretable bounds on the divergence between degree and eigenvector centrality as a function of the strength of assortativity and local structure. These are the first analytical results that link commonly used network statistics to quantitative guarantees on the accuracy of degree-based proxies for eigenvector centrality.
\\
\indent
We are not the first to study how assortativity-inducing degree-preserving rewiring influences the eigenstructure of a network. Some years ago, \citet{VanMieghem2010AssortativitySpectra} used degree-preserving rewiring to systematically tune assortativity and examine how the eigen spectrum responds. Our analysis is close in spirit but differs in two key respects. First, we extend the rewiring-based perspective beyond assortativity to a broader class of local structures. Second, and more importantly, we embed these constructions within a perturbation-theoretic framework by linking the resulting perturbation matrices to the Stewart--Sun eigenvector bound. In this sense, our paper bridges several strands of work in network science and spectral graph theory: rewiring-based control of assortativity, the study of eigenvector localization, and the use of spectral perturbation theory to assess the robustness of centrality measures. It is perhaps worth noting that we take a constructive approach to proof, i.e., we work through algorithms that alter a graph towards desired geometries. This allows us to characterize the angle of deviation for finite-sized graphs with specific spectral properties. 
\\
\indent
The paper is organized as follows. Section~\ref{sec:preliminaries} sets notation and collects the spectral preliminaries needed for our main arguments, including the Stewart--Sun perturbation bound. Section~\ref{sec:bound_moments} bounds the moments of the angle of deviation with the moments of network statistics. Section~\ref{sec:assortativity} details the assortativity-inducing degree-preserving rewiring process. It then applies the moment-bound proposition to assortativity, while noting how the constants in the bound depend on certain structural features of the network. Section~\ref{sec:local_structures} details the degree-preserving rewiring processes used to generate three distinct local structures: communities, core-periphery patterns, and cycles. We generalize each of these structures, with $K$ communities, $k$ cycles, and perhaps most interestingly, fractal core-periphery wherein the periphery itself consists of `miniature' core-periphery structures. We then apply the moment bound proposition to bound the angle of deviation with measures of communities, core-periphery, and cycles. Section \ref{sec:bound_diff} discusses how the heavy tails of the degree distribution shape the bounds on the angle of deviation between the degree-vector and eigenvector. Section~\ref{sec:concluding} offers concluding remarks. Appendix \ref{app:weighted_networks} extends the analytical results to weighted directed graphs by considering the continuous analogue of our discrete rewiring setting.  Appendix \ref{app:simulations} presents numerical simulations that estimate the exact angle of deviation for specific quantums of assortativity and community structures. These simulations support our analytical results about upper bounds on the angle of deviation and further clarify the mechanisms involved in generating these deviations.

\section{Spectral Preliminaries and the Stewart--Sun Perturbation Bound} \label{sec:preliminaries}

\subsection{Basic notation}
$\mathbb{E}$ stands for expectation and $\mathbb{V}$ for variance.  Matrices are denoted by bold capital letters, vectors by bold lowercase letters, and scalars by plain letters: for example, $\mathbf X$ is a matrix, $\mathbf x$ a vector, and $x$ a scalar. We write $\mathbf X^{\top}$ and $\mathbf X^{-1}$ for the transpose and inverse of $\mathbf X$, respectively. For vectors $\mathbf x\in\mathbb R^n$ we use $\|\mathbf x\|_2$ to denote the Euclidean norm. Given a degree sequence $[d_1, d_2, ..., d_n]$, let $\mathbf{d}$ denote the $\ell_2$-normalized degree vector corresponding to this degree sequence. $\mathbf{d}$ may be interpreted as in-degree or out-degree based on context. For any square matrix $\mathbf X$ we write $\lambda_k(\mathbf X)$ for its $k$-th eigenvalue, ordered by nonincreasing absolute value,
\[
|\lambda_1(\mathbf X)| \;\ge\; |\lambda_2(\mathbf X)| \;\ge\; \cdots
\]
and when the matrix is clear from context we abbreviate $\lambda_1(\mathbf X)$ to $\lambda_1$. Given any square matrix $\mathbf X\in\mathbb R^{n\times n}$, we define its spectral norm by 
\[
\|\mathbf X\|_2
\;:=\;
\sup\bigl\{\|\mathbf X\mathbf z\|_2 : \mathbf z\in\mathbb R^n,\ \|\mathbf z\|_2 = 1\bigr\}
\]
$\mathbf v^{\mathrm{R}}$ and $\mathbf v^{\mathrm{L}}$ denote the unit leading
right and left eigenvectors of $\mathbf A$, respectively. For brevity, we write
$\mathbf v$ for whichever of these two unit eigenvectors is relevant in a given
context. The Euclidean norm of a vector $\mathbf x$ is denoted by $||\mathbf x||_2$.

\begin{mydef}[Distance between two vectors as an angle]\label{def:angular-distance}
For two nonzero vectors $\mathbf x,\mathbf y\in\mathbb R^n$ we define their
angular distance as the acute angle between them
\[
\theta_{\mathbf x,\mathbf y}
\;:=\;
\arccos\!\biggl(
\frac{|\langle \mathbf x,\mathbf y\rangle|}
     {\|\mathbf x\|_2\,\|\mathbf y\|_2}
\biggr)
\;\in\;[0,\tfrac{\pi}{2}]
\]
which is the standard notion of angle between vectors in Euclidean space. When $\mathbf x$ and $\mathbf y$ are unit vectors, this reduces to
$\theta_{\mathbf x,\mathbf y}=\arccos(|\langle \mathbf x,\mathbf y\rangle|)$. Whenever we compare or distance between a degree vector and an eigenvector (for example, an alignment $\theta_{\mathbf d,\mathbf v}$) is to be understood as comparing `out-degree and right-eigenvector' or `in-degree and left-eigenvector'.
\enremark
\end{mydef}

\subsection{Neutral matrix and its alignment with the degree vector}

\begin{mydef}[Neutral network]\label{def:neutral-network}
A baseline neutral network is an unweighted directed graph on $n$ nodes
with adjacency matrix $\mathbf A\in\mathbb{R}_{\ge 0}^{n\times n}$.
We assume that $\mathbf A$ is irreducible, that the Perron root $\lambda_1(\mathbf A)$ is
simple, and that $\mathbf A$ is diagonalizable, so that the Perron eigenvectors
are well-defined up to scale.\footnote{We throughout assume that the rewiring steps we consider are
sufficiently few and local that the graph remains irreducible along the
trajectory. Equivalently, one may restrict attention to rewiring moves that
preserve strong connectivity (e.g., by rejecting swaps that break
irreducibility).}
\\
\indent
Neutrality means that the adjacency matrix contains no systematic structure
beyond that implied by its in- and out-degree sequences. More specifically, let $\mathbf d^{\mathrm{out}}$ be an out-degree sequence and $\mathbf d^{\mathrm{in}}$ be an in-degree sequences. Let $m:=\mathbf 1^{\top}\mathbf d^{\mathrm{out}}$ denote the number of directed edges. We take the neutral baseline to be the degree-only (rank-one) matrix
\[
\mathbf A
\;:=\;
\frac{\mathbf d^{\mathrm{out}}(\mathbf d^{\mathrm{in}})^{\top}}{m}
\]
\enremark
\end{mydef}

\begin{proposi}[Degree vector as a proxy for the leading eigenvector in neutral networks]
\label{prop:neutral-degree-evec}
 Let $\mathbf v$ denote the unit Perron eigenvector of $\mathbf A$ ,
and let $\mathbf d$ denote the corresponding unit degree vector. Then
\[
\theta_{\mathbf v,\mathbf d} \;\approx\; 0
\]
so that degree centrality coincides with eigenvector centrality in the neutral
baseline \citep[Sec.~7.8]{Newman2010Networks}. Note that when $\mathbf v $ is the right eigenvector, then $\mathbf d $ is the out-degree, equivalently for the left eigenvector and the in-degree. 
\enremark
\end{proposi}

\subsection{Perturbation matrices and their spectral norms}
\begin{mydef}[Spectral gap]\label{def:spectral-gap}
Let $\mathbf A\in\mathbb R^{n\times n}$ be a matrix with simple eigenvalues $\lambda_1,\dots,\lambda_n$, with $\lambda_1$ as the leading eigenvalue. The spectral gap of $\mathbf A$ is defined as
\[
\gamma(\mathbf A)
\;:=\;
\min_{j\ge 2} \bigl|\lambda_1 - \lambda_j\bigr|
\]
The spectral gap measures how well separated the leading eigenvalue is from the rest of the spectrum.
\enremark
\end{mydef}

\begin{mydef}[Perturbation matrices]
Consider a sequence of perturbation matrices
\[
\boldsymbol{\Delta}^{(1)},\,\boldsymbol{\Delta}^{(2)},\,\dots,\,\boldsymbol{\Delta}^{(t)} \in \mathbb R^{n\times n}
\]
with entries in $\{-1,0,1\}$. Each $\boldsymbol{\Delta}^{(i)}$ represents a single
degree-preserving rewiring step involving two edges and has exactly four nonzero
entries: two equal to $+1$ and two equal to $-1$. We write the cumulative perturbation
after $t$ steps as
\[
\boldsymbol{\Omega}^{(t)} \;:=\; \sum_{i=1}^t \boldsymbol{\Delta}^{(i)}
\]
Accordingly, the adjacency matrix after $t$ perturbations is
\[
\mathbf A^{(t)}
\;:=\;
\mathbf A \;+\; \boldsymbol{\Omega}^{(t)}
\]
We assume that $\mathbf A^{(t)}$ is irreducible and diagonalizable for all $t$ (equivalently, we reject any rewiring that breaks these properties (since $\mathbf A^{(0)}=\mathbf A$ is assumed to be irreducible and diagonalizable).
\enremark
\end{mydef}

\begin{assumption}[$r$-bounded participation rewiring]\label{assump:r-bounded-participation-strong}
Each node participates in at most $r$ rewiring steps. Equivalently, there exist
permutation matrices $\mathbf{\Pi}_1,\dots,\mathbf{\Pi}_r$ and a partition of
$\{1,\dots,k\}$ into $r$ subsets $I_1,\dots,I_r$ such that for each
$s\in\{1,\dots,r\}$
\[
\mathbf{\Pi}_s^{\top}\Big(\sum_{i\in I_s}\boldsymbol{\Delta}^{(i)}\Big)\mathbf{\Pi}_s
\]
is block-diagonal with one block for each $\boldsymbol{\Delta}^{(i)}$, $i\in I_s$.
\enremark
\end{assumption}

\begin{proposi}[Spectral norm bound for the cumulative rewiring perturbation]
\label{prop:delta-omega-norm}
If each vertex participates in at most $r\in\mathbb N$ swaps, then
\[
\|\boldsymbol{\Omega}^{(t)}\|_2 \;\le\; 2r
\]

\begin{proof}
Fix a vertex $u$ and let $s(u)$ be the number of swaps in which $u$ participates.
In any single degree-preserving swap, a participating vertex has exactly two
incident edges whose adjacency entries flip (one $1\!\to\!0$ and one $0\!\to\!1$).
Consequently, in the cumulative matrix $\boldsymbol{\Omega}^{(t)}$, each time $u$
participates it can contribute at most two nonzero entries of magnitude $1$ in
row $u$, and likewise at most two such entries in column $u$. Hence
\[
\sum_{j}\bigl|\Omega^{(t)}_{uj}\bigr|\;\le\;2s(u),
\qquad
\sum_{i}\bigl|\Omega^{(t)}_{iu}\bigr|\;\le\;2s(u)
\]
Let $s_{\max}:=\max_{u}s(u)\le r$. Taking maxima over $u$ gives the induced
$\ell^1$- and $\ell^\infty$-operator norms:
\[
\|\boldsymbol{\Omega}^{(t)}\|_1
=\max_{j}\sum_{i}\bigl|\Omega^{(t)}_{ij}\bigr|
\;\le\;2s_{\max},
\qquad
\|\boldsymbol{\Omega}^{(t)}\|_\infty
=\max_{i}\sum_{j}\bigl|\Omega^{(t)}_{ij}\bigr|
\;\le\;2s_{\max}
\]
Finally, the standard inequality between induced norms yields
\[
\|\boldsymbol{\Omega}^{(t)}\|_2
\;\le\;
\sqrt{\|\boldsymbol{\Omega}^{(t)}\|_1\,\|\boldsymbol{\Omega}^{(t)}\|_\infty}
\;\le\;
\sqrt{(2s_{\max})(2s_{\max})}
\;=\;
2s_{\max}
\;\le\;
2r
\]
which proves the claim.
\end{proof}
\end{proposi}

\subsection{Stewart--Sun eigenvector perturbation bound}

\begin{mydef}[Distortion factor]\label{def:alignment}
Let $\mathbf A\in\mathbb R^{n\times n}$ be a diagonalizable matrix with
eigendecomposition
\[
\mathbf A \;=\; \mathbf V \boldsymbol{\Lambda}\,\mathbf V^{-1}
\]
where $\boldsymbol{\Lambda}$ is diagonal with the eigenvalues of $\mathbf A$ on
its diagonal, and $\mathbf V=[\mathbf v_1,\dots,\mathbf v_n]$ collects the
corresponding (right) eigenvectors as columns. The distortion factor of $\mathbf A$ is defined by
\[
\kappa(\mathbf A) \;:=\; \|\mathbf V\|_2 \,\|\mathbf V^{-1}\|_2
\]
When the underlying matrix is clear from context, we write $\kappa$ for
$\kappa(\mathbf A)$. This quantity measures how far the eigenvector matrix
$\mathbf V$ is from being orthogonal\footnote{For directed networks, adjacency matrices need not admit an orthogonal eigenbasis. Here ``normal'' means that $\mathbf A^\top \mathbf A=\mathbf A \mathbf A^\top$, which is the relevant benchmark for $\kappa(\mathbf A)=1$. For a $\{0,1\}$ adjacency matrix $\mathbf A$, one has $(\mathbf A^\top\mathbf A)_{ii}=d_i^{\mathrm{in}}$ and $(\mathbf A\mathbf A^\top)_{ii}=d_i^{\mathrm{out}}$. Hence normality implies $d_i^{\mathrm{in}}=d_i^{\mathrm{out}}$ for every node $i$. Therefore any node-level divergence $d_i^{\mathrm{in}}\neq d_i^{\mathrm{out}}$ rules out $\kappa(\mathbf A)=1$, although degree balance alone does not guarantee normality.}

: in particular, $\kappa(\mathbf A)=1$ when
$\mathbf V$ is orthogonal, i.e.,
\[
\mathbf V^\top \mathbf V=\mathbf I
\quad\Longleftrightarrow\quad
\langle \mathbf v_i,\mathbf v_j\rangle = 0\ \ \text{for all }i\neq j
\ \ \text{and}\ \ \|\mathbf v_i\|_2=1\ \ \text{for all }i
\]
Note that $\kappa=1$ if and only if the eigenvectors are pairwise orthogonal\footnote{A somewhat rough interpretation of near pair-wise orthogonality is that the nodes that are most significant in generating one type of structure are disjoint from those that generate another, and that this is true for all structures. This would mean that, for example, the nodes most responsible for heavy tails in degree distribution have little role to play in triangles.}. 
\enremark
\end{mydef}

\begin{result}[Stewart--Sun eigenvector perturbation bound]\label{res:stewart-sun}
Let $\mathbf A\in\mathbb R^{n\times n}$ be diagonalizable with simple eigenvalues.
Let $\gamma=\gamma(\mathbf A)$ be its spectral gap (Def~\ref{def:spectral-gap})
and $\kappa=\kappa(\mathbf A)$ its distortion factor
(Def~\ref{def:alignment}). $\mathbf v$ is the unit leading eigenvector of $\mathbf A$ and  $\mathbf v^{(t)}$ the unit leading eigenvector of $\mathbf A^{(t)}$. Let
$\theta_{\mathbf v,\mathbf v^{(t)}}$ denote the acute angle between
$\mathbf v$ and $\mathbf v^{(t)}$. If
\[
\|\boldsymbol{\Omega}^{(t)}\|_2 \;<\; \frac{\gamma}{\kappa}
\]
then the \citet{StewartSun1990MatrixPerturbation} perturbation bound gives
\[
\label{eq:stewart-sun-nonsym}
\sin \theta_{\mathbf v,\mathbf v^{(t)}}
\;\le\;
\frac{\kappa\,\|\boldsymbol{\Omega}^{(t)}\|_2}{\gamma}
\]
\enremark
\end{result}
The bound formalizes the idea the perturbation $\boldsymbol{\Omega}^{(t)}$ can only rotate the leading eigenvector by a small angle if the leading eigenvalue is well separated from the rest of the spectrum (large $\gamma$) and if the eigenbasis is not too ill-conditioned (small $\kappa$). In essence, $\kappa$ and (reciprocal of) $\gamma$ capture information about how petrubations to smaller eigenmodes translate to a perturbation to the leading eigenmode because of the dependencies between them. More specifically, a small spectral gap (small $\gamma$) means the top modes have nearly equal growth rates, so even perturbations that primarily affect a lower mode can readily mix with the leading mode. A large distortion factor (large $\kappa$) means the eigenvectors are far from orthogonal, so the eigenmodes are geometrically entangled: a perturbation expressed in a `lower' direction can leak into the leading direction through the ill-conditioned change-of-basis. Thus $\gamma$ captures dynamic separability of modes, while $\kappa$ captures geometric separability. Naturally, either form of coupling allows disturbances to lower-ranked structures to indirectly move the leading eigenmode.

\begin{proposi}[Angle of deviation between the degree vector and the eigenvector]\label{prop:deg-evec-SS}
Let $\mathbf A$ be neutral (Def.~\ref{def:neutral-network}), with unit degree
vector $\mathbf d$ and unit leading eigenvector $\mathbf v$. Let $\mathbf v^{(t)}$
be the unit leading eigenvector of $\mathbf A^{(t)}=\mathbf A+\boldsymbol{\Omega}^{(t)}$.
Assume the Stewart--Sun condition
\[
\|\boldsymbol{\Omega}^{(t)}\|_2 \;<\; \frac{\gamma}{\kappa}
\]
where $\gamma=\gamma(\mathbf A)$ and $\kappa=\kappa(\mathbf A)$ are as in
Result~\ref{res:stewart-sun}. Then
\[
\theta_{\mathbf d,\mathbf v^{(t)}}
\;\le\;
\theta_{\mathbf d,\mathbf v}
\;+\;
\arcsin \Bigl(\tfrac{\kappa\,\|\boldsymbol{\Omega}^{(t)}\|_2}{\gamma}\Bigr)
\]
If, moreover, each node participates in at most $r$ swaps so that
$\|\boldsymbol{\Omega}^{(t)}\|_2\le 2r$ (Prop.~\ref{prop:delta-omega-norm}), then whenever
$\tfrac{2\kappa r}{\gamma}<1$
\[
\theta_{\mathbf d,\mathbf v^{(t)}}
\;\le\;
\theta_{\mathbf d,\mathbf v}
\;+\;
\arcsin \Bigl(\tfrac{2\kappa r}{\gamma}\Bigr)
\]
Since $\theta_{\mathbf d,\mathbf v}\approx 0$ (Prop.~\ref{prop:neutral-degree-evec}), we have
\[
\sin\theta_{\mathbf d,\mathbf v^{(t)}} \;\lesssim \; \frac{2\kappa r}{\gamma}
\]
\enremark
\end{proposi}
In other words, given our degree-preserving rewiring process, the angle of deviation between degree vector and the eigenvector is bounded from above by the maximum number of rewirings per node ($r$), a measure of the orthogonality of the eigenmodes ($\kappa$), and the reciprocal of the spectral gap ($\lambda$).

\section{Bounding moments of angle of deviation with moments of network statistic} \label{sec:bound_moments}

\begin{conjecture}[Wandering angle under strictly $\phi$-improving rewiring]
\label{conj:wandering_angle}
Fix in- and out-degree sequences and let $\mathcal S$ be the (finite) set of directed
adjacency matrices reachable from an initial $\mathbf A^{(0)}$ by degree-preserving
single-swap rewirings. Fix a unit degree vector $\mathbf d$ that is invariant on
$\mathcal S$.

Consider a degree-preserving statistic-driven rewiring process that pathwise increases a chosen statistic $\phi$
pertaining to assortativity and/or local or mesoscopic structure (e.g.\ triangles,
$k$-cycles, communities, core--periphery). Thus, along the realized trajectory
$\{\mathbf A^{(t)}\}_{t\ge 0}\subset \mathcal S$ of accepted swaps,
\[
\phi(\mathbf A^{(t+1)}) \;\geq\; \phi(\mathbf A^{(t)}) \qquad \text{for all } t\ge 0.
\]
For each $\mathbf A^{(t)}\in\mathcal S$, let $\mathbf v(\mathbf A^{(t)})$ be the unit
leading eigenvector and let $\theta_t$ be the angle of deviation between
$\mathbf v(\mathbf A^{(t)})$ and the fixed degree vector $\mathbf d$.
\\
\indent
The conjecture is that, for a broad class of statistics $\phi$ that reward local or
mesoscopic structure, the induced angle process $(\theta_t)_{t\ge 0}$ typically does
not drift with a fixed sign. That is, although $\phi(\mathbf A^{(t)})$ increases
with $t$, the sequence $\theta_t$ may increase over some epochs and decrease over
others. In particular, $\theta_t$ need not be monotone in $\phi(\mathbf A^{(t)})$
either pathwise or in expectation. In essence, the angle of deviation between the
degree vector and the eigenvector can `wander'. The mechanism that generates this
wandering is the tension between the two opposing forces induced by the rewiring
procedure:
\begin{enumerate}
\item[\textup{(i)}] Pull away from degree: conditioning on
$\phi(\mathbf A^{t+1})\geq\phi(\mathbf A^{t})$ biases accepted swaps toward edge
rearrangements that create the local patterns rewarded by $\phi$. These patterns
reallocate Perron mass across vertices without changing degrees, pushing the eigenvector
$\mathbf v(\mathbf A^{(t)})$ away from the fixed degree vector $\mathbf d$, thereby 
increasing $\theta_t$.

\item[\textup{(ii)}] Pull toward degree: the collection of $\phi$-improving swaps is
biased towards high-degree nodes. From a combinatorial point of view, there is a
greater likelihood of candidate endpoints for cycles, communities, and other local
structures being at or near high-degree nodes. Hence, even under uniform proposals,
accepted swaps may disproportionately endow high-degree nodes with additional local
structure\footnote{Note that the condition in Steward--Sun that each node can
participate in at most $r$ rewiring limits how much local structure can accumulate
around any single vertex along the rewiring trajectory. The bounded participation
condition, therefore, limits the `opportunity effect' when $r$ is sufficiently small
compared to the heaviness of the tails of the degree distribution.}. Since
$\mathbf v(\mathbf A)$ frequently correlates with degree in heterogeneous graphs,
this reinforcement can pull $\mathbf v(\mathbf A^{(t)})$ back toward $\mathbf d$ and
thereby decrease $\theta_t$.
\end{enumerate}
The interplay of \textup{(i)} and \textup{(ii)} can therefore produce a trajectory
for $\theta_t$ that wanders within $[0,\pi]$ despite the strict monotonicity of
$\phi(\mathbf A^{(t)})$ along each step\footnote{This conjecture helps explain why the empirical evidence on the matter is mixed. One line of work shows that as local or mesoscopic structure accumulates, the leading eigenvector may concentrate on a small set of vertices and drift away from degree-based rankings \citep{Sharkey2019CutVertex,PastorSatorrasCastellano2016}. But the opposite tendency can also occur: within a fixed degree sequence, there are neutral realizations in which degree and eigenvector centrality remain close. This includes maximum-entropy/BFD graphs \citep{AtayBiyikogluGraphEntropy}. Some rewiring protocols effectively add structure in ways that track degree rather than compete with it. In short, different networks and protocols can therefore generate different outcomes. Our conjecture gives reasons for why the empirical record does not point in a single direction.}
\\
\indent
(In Def~\ref{def:rewire}, we impose some structure upon the relation between the
evolution of the network statistic and the angle of deviation between the degree
vector and eigenvector).
\enremark
\end{conjecture}

\begin{mydef}[Statistic-driven rewiring with mean-angle admissibility]
\label{def:rewire}
Let $\phi:\mathcal S\to\mathbb R$ be a network statistic and let
$\mathbf A^{(0)}\in\mathcal S$ be a neutral network (Def.~\ref{def:neutral-network}).
We index the process by the number of accepted perturbations $t$. Given $\mathbf A^{(t)}$, let $\mathcal N(\mathbf A^{(t)})\subseteq\mathcal S$ denote its (degree-preserving) swap neighborhood and define the $\phi$-upper-contour set
\[
\mathcal B^{(t)}
\;:=\;
\bigl\{\mathbf B\in \mathcal N(\mathbf A^{(t)}):\ \phi(\mathbf B)\ge \phi(\mathbf A^{(t)})\bigr\}
\]
Assume $\mathcal B_t\neq\emptyset$ along the horizon of interest\footnote{We work in regimes
with only a limited amount of assortativity tuning and a modest accumulation of local structures, far from saturation, so the $\phi$-upper contour remains nontrivial along the realized trajectory.}. Let $\widetilde{\mathcal B}_t\subseteq \mathcal B_t$ be a subset satisfying the
mean-angle admissibility condition
\[
\mathbb E \bigl[\theta(\mathbf A^{(t+1)})\mid \mathbf A^{(t)}\bigr]\ \ge\ \theta(\mathbf A^{(t)}),
\qquad \text{with } \mathbf A^{(t+1)}\in \widetilde{\mathcal B}^{(t)}
\]
In general, there may be many such admissible subsets. We take $\widetilde{\mathcal B}^{t}$ to be
the largest (by inclusion) subset of $\mathcal B^{t}$ for which the above inequality holds.\footnote{
This is weaker than requiring $\theta(\mathbf A^{(t+1)})\ge \theta(\mathbf A^{(t)})$ for every accepted move. Imposing only nonnegative drift in expectation typically makes the admissible set much less likely to be empty when $\mathcal B_t$ is large.} Finally, choose $\mathbf A^{(t+1)}\in \widetilde{\mathcal B}_t$ according to a  sampling rule on $\widetilde{\mathcal B}_t$. Define $\phi_t:=\phi(\mathbf A^{(t)})$. By construction, $\phi_{t+1}\ge \phi_t$ along every realized trajectory and $\theta_t$ is a submartingale.
\enremark
\end{mydef}

Note that the angle-biased selection rule above is a deliberately conservative modeling choice: we explicitly allow the degree--eigenvector angle to deviate as the rewiring procedure builds up assortativity or other local structures. Consequently, the bounds we derive in later sections on how much $\theta(\mathbf A)$ can change as a function of assortativity (or other structure-rewarding statistics) are upper bounds: they quantify the maximal misalignment compatible with the imposed statistic trajectory and the rewiring budget, not a typical or inevitable misalignment along every realization.

\begin{proposi}[Existence and monotonicity of moments for a bounded network statistic]
\label{prop:stat_moments}
Let $\{\mathbf A^{(t)}\}_{t\ge 0}$ be the network statistic driven degree-preserving
rewiring process from Def~\ref{def:rewire}. Assume that $\phi$ is bounded on the state space reachable by the rewiring, i.e., there exist constants $\underline\phi<\overline\phi$ such that
\[
\underline\phi \le \phi(\mathbf A)\le \overline\phi
\qquad\text{for all reachable }\mathbf A
\]
Then for every $t\ge 0$ and every $p\ge 1$ the moment $\mathbb E[|\phi_t|^p]$ is
well-defined and finite. In particular, $\mathbb E[\phi_t]$ and $\mathbb E[\phi_t^2]$
exist for all $t$.

If, in addition, $\phi$ is strictly positive on the reachable state space
(e.g., $0<\phi(\mathbf A)\le \overline\phi$), then for every $p\ge 1$ the $p$th
moment is  increasing in $t$:
\[
\mathbb E[\phi_{t+1}^p]\;\geq\;\mathbb E[\phi_t^p]
\qquad\text{for all }t\ge 0
\]
Indeed, $t$ counts accepted swaps, so the acceptance rule enforces the pathwise
increase $\phi_{t+1}\geq\phi_t$ almost surely. Under positivity, the map
$x\mapsto x^p$ is increasing on $\mathbb R_{>0}$, hence
$\phi_{t+1}^p\geq\phi_t^p$ almost surely, and taking expectations yields the claim.
In particular,
\[
\mathbb E[\phi_{t+1}] \; \geq \; \mathbb E[\phi_t],
\qquad
\mathbb E[\phi_{t+1}^2] \;\geq\; \mathbb E[\phi_t^2],
\qquad\text{for all }t\ge 0
\]

Note that even under positivity and  pathwise increase, the variance $\mathbb{V}(\phi_t)$ need not be monotone in $t$: while both $\mathbb E[\phi_t]$ and $\mathbb E[\phi_t^2]$ increase, the difference $\mathbb{V}(\phi_t)=\mathbb E[\phi_t^2]-\bigl(\mathbb E[\phi_t]\bigr)^2$ can increase or decrease depending on their relative rates of growth. In fact, in general, $\mathbb{V}(\phi_t)$ will tend to increase in the early stages of rewiring as more and diverse networks become `reachable' within one step. After a sufficiently large number of rewirings, $\mathbb{V}(\phi_t)$ will tend to decrease as the statistic reaches the neighborhood of its upper bound, and therefore networks reachable within one step tend not to be able to offer much of an improvement. 
\enremark
\end{proposi}

\begin{proposi}[Moment transfer via local linearization]
\label{prop:moment_bounds_angle_from_stat}
Let $\{\mathbf A^{(t)}\}_{t\ge0}$ be an $\mathcal S$-valued rewiring process with
\[
\mathbf A^{(t)}=\mathbf A^{(0)}+\boldsymbol{\Omega}^{(t)},
\qquad 
\boldsymbol{\Omega}^{(t)}=\sum_{s=0}^{t-1}\boldsymbol{\Delta}^{(s)}
\]
where each swap induces a low-rank perturbation $\boldsymbol{\Delta}^{(s)}$
with uniformly bounded size (so the steps are marginal in operator norm). Let
\[
\theta_t:=\theta(\mathbf A^{(t)})\in[0,\pi/2],
\qquad
\phi_t:=\phi(\mathbf A^{(t)})
\]
for a network statistic $\phi:\mathcal S\to\mathbb R$.

Assume that over the horizon of interest the trajectory remains in a subset
$\mathcal S_0\subseteq\mathcal S$ on which the leading eigenvalue is
well separated and the eigenbasis is not too ill-conditioned. In particular,
on any reachable set $\mathcal S_0$ supporting the affine comparison below one
may take uniform spectral parameters\footnote{The reachable region $\mathcal S_0$ (hence $\kappa_*,\gamma_*$) may depend on the statistic $\phi$ and the acceptance rule, since different driving statistics can steer the trajectory through regions with different spectral conditioning.}
\[
\kappa_*:=\sup_{\mathbf A\in\mathcal S_0}\kappa(\mathbf A),
\qquad
\gamma_*:=\inf_{\mathbf A\in\mathcal S_0}\gamma(\mathbf A)\;>\;0
\]

On such a region, eigenvector perturbation theory (Stewart--Sun) implies that
$\mathbf A\mapsto \theta(\mathbf A)$ is locally Lipschitz in the operator norm:
there exists $L_\theta>0$ such that for all $\mathbf A,\mathbf B\in\mathcal S_0$,
\[
|\theta(\mathbf B)-\theta(\mathbf A)|
\;\le\;
L_\theta\,\|\mathbf B-\mathbf A\|_2,
\qquad\text{with } L_\theta \asymp \kappa_*/\gamma_*
\]

Assume further that, along $\mathcal S_0$, the statistic $\phi$ tracks the same
marginal edits in the sense that there exist constants $a^\pm\ge 0$,
$b^\pm\in\mathbb R$, and a finite-size slack $\varepsilon_n\ge0$ such that the
following affine bounds hold:
\[
\label{eq:affine_sandwich_phi_rewrite}
a^{-}\,\phi(\mathbf A)+b^{-}-\varepsilon_n
\ \le\
\theta(\mathbf A)
\ \le\
a^{+}\,\phi(\mathbf A)+b^{+}+\varepsilon_n,
\qquad \forall\,\mathbf A\in\mathcal S_0
\]
Here the effective constants $(a^\pm,b^\pm,\varepsilon_n)$ may depend on the
uniform spectral controls $(\kappa_*,\gamma_*)$ (through the stability of
$\theta(\cdot)$ under marginal edits) and on statistic-specific parameters that
govern how $\phi(\cdot)$ responds to an accepted swap (e.g.\ partition
granularity, $d_{\max}$-type caps, motif size, or acceptance-rule constraints).
\\
\indent
If $\phi$ is bounded on the reachable state space (so that $\mathbb{E}[\phi_t]$ and
$\mathbb{E}[\phi_t^2]$ exist), then for every $t$
\[
a^{-}\,\mathbb{E}[\phi_t]+b^{-}-\varepsilon_n
\ \le\ \mathbb{E}[\theta_t]\ \le\
a^{+}\,\mathbb{E}[\phi_t]+b^{+}+\varepsilon_n
\]
Moreover, with $c^{+}:=|b^{+}|+\varepsilon_n$
\[
\mathbb{E}[\theta_t^2]\ \le\
(a^{+})^2\,\mathbb{E}[\phi_t^2]
\;+\;2a^{+}c^{+}\,\mathbb{E}[|\phi_t|]
\;+\;(c^{+})^2
\]
\enremark
\end{proposi}

\begin{corollary}[Multiplicative envelope for signed statistics]
\label{cor:multiplicative_envelope_abs}
Fix a reachable region $S_0$ and a statistic $\phi:\mathcal S\to\mathbb R$. Suppose the affine upper comparison from Proposition~5 holds on $S_0$:
\[
\theta(A)\;\le\; a_\phi^+\,\phi(A)+b_\phi^+ + \varepsilon_n,
\qquad \forall A\in S_0
\]
Then, for all $A\in S_0$,
\[
\theta(A)\;\le\; a_\phi^+\,|\phi(A)|+\underbrace{\bigl(|b_\phi^+|+\varepsilon_n\bigr)}_{=:c_\phi}
\]
Let $M_\phi:=\max\{a_\phi^+,c_\phi\}$. Since $a_\phi^+|\phi(A)|+c_\phi\le M_\phi|\phi(A)|+M_\phi$,
we obtain the purely algebraic multiplicative envelope
\[
\theta(A)\;\le\; M_\phi\bigl(1+|\phi(A)|\bigr),
\qquad \forall A\in S_0
\]
and hence along the trajectory
\[
\mathbb E[\theta_t]\;\le\; M_\phi\Bigl(1+\mathbb E\bigl[|\phi_t|\bigr]\Bigr)
\]
\medskip
\noindent
\medskip
\noindent
\textit{(Chaining linear bounds to obtain a product envelope for $M_\phi$.)}
Assume that on $S_0$ the following \emph{linear} controls hold:

\smallskip
\noindent
(i) \emph{Linear spectral sensitivity.}
There is a constant $L_\theta\lesssim \kappa_*/\gamma_*$ such that for all $A\in S_0$,
\[
\theta(A)\;\le\; \theta(A^{(0)}) + L_\theta\,\|A-A^{(0)}\|_2 
\]

\smallskip
\noindent
(ii) \emph{Participation-controlled perturbation size.}
Under $r_\phi$-bounded participation, the cumulative perturbation
$\Omega^{(t)}=\sum_{s=0}^{t-1}\Delta^{(s)}$ satisfies
\[
\|\Omega^{(t)}\|_2 \;\le\; \sqrt{\|\Omega^{(t)}\|_1\,\|\Omega^{(t)}\|_\infty}
\;\le\; c_\Omega\, r_\phi 
\]
where $\|\cdot\|_\infty$ and $\|\cdot\|_1$ are the maximum absolute row- and column-sum norms. Here $c_\Omega>0$ is universal (e.g.\ $c_\Omega=2$ under the two-edge swap model), because $r_\phi$-bounded participation bounds the total signed mass that can accumulate in any fixed row or column of $\Omega^{(t)}$.

\smallskip
\noindent
(iii) \emph{Linear leverage of the driving statistic.}
There is a statistic-specific scale $\Lambda_\phi$ such that each accepted swap changes the statistic
by at most $c_\phi^{\mathrm{lev}}\Lambda_\phi$:
\[
|\phi_{t+1}-\phi_t|\;\le\;c_\phi^{\mathrm{lev}}\Lambda_\phi
\]
with $c_\phi^{\mathrm{lev}}$ universal.\footnote{As before, this is a bounded-differences property: a single swap alters only $O(1)$ edge incidences, so $|\Delta\phi|$ is controlled by degree caps, partition granularity, motif length, and the normalization of $\phi$. In normalized statistics, $\Lambda_\phi$ typically already carries factors such as $1/m$, $1/|C_k|^2$, etc., which prevents spurious growth with $n$.}

\smallskip
\noindent
Chaining (i) and (ii) gives the basic linear envelope
\[
\theta(A^{(t)})\;\le\;\theta(A^{(0)}) + \Bigl(\frac{\kappa_*}{\gamma_*}\Bigr)\,(c_\Omega r_\phi)
\]
Combining this with (iii) (to express the same perturbation budget in the natural leverage scale of the driving statistic) yields the product-form envelope\footnote{Whenever the ingredients are linear in their respective controls, chaining them
automatically produces a multiplicative envelope: if $u\le c_1 x$, $x\le c_2 y$, and $y\le c_3 z$
(with nonnegative quantities), then substitution gives $u\le (c_1c_2c_3)\,z$. In the present
setting one may read the chain schematically as
\[
\theta-\theta_0 \;\le\; L_\theta\,\|\Omega\|_2,
\qquad
L_\theta \;\lesssim\; \frac{\kappa_*}{\gamma_*},
\qquad
\|\Omega\|_2 \;\le\; c_\Omega\, r_\phi,
\]
together with the statistic-side leverage scale $\Lambda_\phi$ that governs the affine comparison
constants (via degree caps, partition granularity, motif length, and normalization). Chaining these
linear bounds yields a product of the contributing factors, while additive intercepts and finite-size
slack are absorbed into the prefactor $C_\phi$.}
\[
M_\phi \;\lesssim\; C_\phi \,\frac{\kappa_*}{\gamma_*}\, r_\phi \,\Lambda_\phi
\]
where $C_\phi$ absorbs baseline terms ($\theta(A^{(0)})$, $|\phi_0|$), universal constants, the normalization conventions of $\phi$, and any residual slack from the affine comparison\footnote{The product form is a bookkeeping consequence of chaining linear inequalities: each step contributes a linear factor (spectral sensitivity, perturbation size, statistic leverage), and their product dominates the final upper envelope once constants are taken large enough to absorb additive slack.}.

\end{corollary}

\section{Assortativity} \label{sec:assortativity}

\begin{mydef}[Newman's assortativity coefficient]
\label{def:neumann}
Let $\mathbf A$ be the adjacency matrix of a (possibly directed) graph with edge set
$\mathcal E(\mathbf A)$ and $m:=|\mathcal E(\mathbf A)|$.  For each vertex $i$, let
$d_i^{\mathrm{in}}$ and $d_i^{\mathrm{out}}$ denote its in- and out-degree. Fix
$(p,q)\in\{\mathrm{in},\mathrm{out}\}^2$.  For each directed edge $(i,j)\in\mathcal E(\mathbf A)$
define the tail and head degrees
\[
x_{ij}:=d_i^{p},\qquad y_{ij}:=d_j^{q}.
\]
Let $\mu_T:=\frac1m\sum_{(i,j)\in\mathcal E(\mathbf A)}x_{ij}$ and
$\mu_H:=\frac1m\sum_{(i,j)\in\mathcal E(\mathbf A)}y_{ij}$, and let
\[
\sigma_T^2:=\frac1m\sum_{(i,j)\in\mathcal E(\mathbf A)}(x_{ij}-\mu_T)^2,\qquad
\sigma_H^2:=\frac1m\sum_{(i,j)\in\mathcal E(\mathbf A)}(y_{ij}-\mu_H)^2
\]
Newman's assortativity coefficient for the choice $(p,q)$ is the Pearson correlation
of $(x_{ij},y_{ij})$ over edges
\[
\phi_{p,q}(\mathbf A)
\;:=\;
\frac{\frac1m\sum_{(i,j)\in\mathcal E(\mathbf A)}(x_{ij}-\mu_T)(y_{ij}-\mu_H)}{\sigma_T\sigma_H}
\]
\enremark
\end{mydef}

\begin{proposi}[Change in Newman's assortativity under degree-preserving rewiring]
\label{prop:assort_rewiring}
Fix $(p,q)\in\{\mathrm{in},\mathrm{out}\}^2$ and let $\phi_{p,q}(\cdot)$ denote
Newman's assortativity coefficient from Def~\ref{def:neumann}.  Let $\mathbf A^{(0)}$ be a
directed adjacency matrix with fixed in- and out-degree sequences, and let $\mathbf A^{(t)}$ be
any matrix reachable from $\mathbf A^{(0)}$ by degree-preserving edge swaps (so $\mathbf A^{(t)}\in\mathcal S$
in the notation of Conjecture~\ref{conj:wandering_angle}).  Then the edge-averaged means and
variances entering $\phi_{p,q}$ are invariants of the degree class:
\[
\mu_T(\mathbf A^{(t)})=\mu_T(\mathbf A^{(0)}),\qquad
\mu_H(\mathbf A^{(t)})=\mu_H(\mathbf A^{(0)})
\]
\[
\sigma_T(\mathbf A^{(t)})=\sigma_T(\mathbf A^{(0)}),\qquad
\sigma_H(\mathbf A^{(t)})=\sigma_H(\mathbf A^{(0)})
\]
and hence the normalization factor
\[
\nu_{p,q}(\mathbf A^{(t)})=\frac{1}{\sigma_T(\mathbf A^{(t)})\,\sigma_H(\mathbf A^{(t)})}
\]
is constant along the rewiring trajectory. Denote this constant by $\nu_{p,q}$. Consequently, the change in assortativity along any degree-preserving rewiring is entirely driven by the change in the edge-wise degree-product sum
\[
\mathcal S_{p,q}(\mathbf A)
\;:=\;
\sum_{(i,j)\in\mathcal E(\mathbf A)} d_i^p\, d_j^q
\]
namely,
\[
\phi_{p,q}(\mathbf A^{(t)})-\phi_{p,q}(\mathbf A^{(0)})
\;=\;
\nu_{p,q}\left[
\frac{1}{m}\,\mathcal S_{p,q}(\mathbf A^{(t)})
\;-\;
\frac{1}{m}\,\mathcal S_{p,q}(\mathbf A^{(0)})
\right]
\]
with $m:=|\mathcal E(\mathbf A^{(0)})|$. In particular, degree-preserving rewiring leaves unchanged the multiset of tail degrees
$\{d_i^p:(i,j)\in\mathcal E(\mathbf A)\}$ and head degrees
$\{d_j^q:(i,j)\in\mathcal E(\mathbf A)\}$. It only changes how tails and heads are paired
across edges. The constant $\nu_{p,q}$ therefore acts as a fixed conversion factor from
changes in the average edge-wise product $\mathcal S_{p,q}(\mathbf A)/m$ to changes in
Newman's assortativity coefficient.
\enremark
\end{proposi}

Note that the conversion factor depends on degree heterogeneity. Recall $\nu_{p,q}=1/(\sigma_T\sigma_H)$, where $\sigma_T^2$ and $\sigma_H^2$ are the
edge-averaged variances of tail and head degrees used in Def~\ref{def:neumann}.  Thus
$\nu_{p,q}$ is large when either $\sigma_T$ or $\sigma_H$ is small, and it is small only
when both $\sigma_T$ and $\sigma_H$ are large. Intuitively, if one side is nearly homogeneous
(say $\sigma_H\approx 0$) while the other is heterogeneous, then small changes in how the
heterogeneous degrees are matched can produce comparatively large changes in the Pearson
correlation, so assortativity moves quickly. By contrast, when both ends of edges are highly
heterogeneous (large $\sigma_T$ and large $\sigma_H$), the background variability is large and a
fixed change in the mean product $\mathcal S_{p,q}(\mathbf A)/m$ translates into a
smaller change in $\phi_{p,q}$.

\paragraph{Assortativity inducing rewiring\\}

Fix $(p,q)\in\{\mathrm{in},\mathrm{out}\}^2$ and let $\phi:=\phi_{p,q}$ be Newman's
assortativity coefficient (Def.~\ref{def:neumann}). Consider the statistic-driven
rewiring rule of Def.~\ref{def:rewire} with this choice of $\phi$. Along the induced
trajectory $\{\mathbf A^{(t)}\}_{t\ge 0}$, a candidate move from $\mathbf A^{(t)}$ to
$\mathbf A^{(t+1)}=\mathbf A^{(t)}+\boldsymbol{\Delta}^{(t)}$ is a degree-preserving
edge swap (Def.~\ref{def:rewire}), removing $(a,b)$ and $(c,d)$ and adding $(a,d)$
and $(c,b)$ (with the usual feasibility conditions). Define the degree-product sum
\[
\mathcal S_{p,q}(\mathbf A)\;:=\;\sum_{(i,j)\in\mathcal E(\mathbf A)} d_i^{p}\,d_j^{q}
\]
For the above swap one has
\[
\mathcal S_{p,q}(\mathbf A^{(t+1)})-\mathcal S_{p,q}(\mathbf A^{(t)})
\;=\;
d_a^{p} d_d^{q} + d_c^{p} d_b^{q} - d_a^{p} d_b^{q} - d_c^{p} d_d^{q}
\]
Moreover, along any degree-preserving trajectory the normalization in $\phi_{p,q}$
is invariant (Prop.~\ref{prop:assort_rewiring}), so
\[
\phi_{p,q}(\mathbf A^{(t+1)})-\phi_{p,q}(\mathbf A^{(t)})
\quad\text{has the same sign as}\quad
\mathcal S_{p,q}(\mathbf A^{(t+1)})-\mathcal S_{p,q}(\mathbf A^{(t)}).
\]
Accordingly, a swap at time $t$ is called positive-assortativity inducing if
$\mathcal S_{p,q}(\mathbf A^{(t+1)})>\mathcal S_{p,q}(\mathbf A^{(t)})$, and
negative-assortativity inducing if the inequality is reversed. Note that since
our assortativity moves are degree-preserving, Proposition~\ref{prop:deg-evec-SS}
applies verbatim: under $r$-bounded participation and a neutral baseline,
$\sin\theta_{\mathbf d,\mathbf v^{(t)}}\le 2\kappa r/\gamma$ (for $2\kappa r/\gamma<1$),
so degree remains a good proxy as long as $r\ll \gamma/\kappa$\footnote{Instead of
bounding each node's participation by $r$, one may bound rewiring at the level of
groups: partition vertices into groups of size $m$ and require $\sum_{u\in G_\ell}s(u)\le R$
for each group. Since then $s_{\max}\le R$, we have $\|\Omega^{(t)}\|_2\le 2R$ and hence
$\sin\theta_{\mathbf d,\mathbf v^{(t)}}\le 2\kappa R/\gamma$ (when $2\kappa R/\gamma<1$).
Writing $R=rm$ shows the bound scales with group size: larger $m$ weakens the worst-case
guarantee because a fixed group budget can be concentrated on fewer nodes.}.

\begin{remark}[Moment bound for the angle via assortativity]\label{rem:moments-angle-assort}
Let $\phi_{p,q}(\mathbf A)$ be the Newman assortativity statistic from
Def~\ref{def:neumann}, $(p,q)\in\{\mathrm{in},\mathrm{out}\}^2$, and along the
degree-preserving assortativity rewiring trajectory set
\[
\phi_t:=\phi_{p,q}(\mathbf A^{(t)}),\qquad \theta_t:=\theta(\mathbf A^{(t)})
\]
The rule is $\phi_{p,q}$-upper-contour (accepted swaps satisfy $\phi_{t+1}\ge \phi_t$). If the
affine upper comparison assumed in Proposition~\ref{prop:moment_bounds_angle_from_stat} holds on
the reachable set for $\phi=\phi_{p,q}$, then Corollary~\ref{cor:multiplicative_envelope_abs}
implies the one-moment control
\[
\mathbb E[\theta_t]\;\le\; M_{p,q}\Bigl(1+\mathbb E\bigl[|\phi_t|\bigr]\Bigr)
\]
for a uniform constant $M_{p,q}$.

\medskip
\noindent\textit{(Assortativity-specific scaling of $M_{p,q}$.)}
Recall that $\phi_{p,q}$ is a normalized covariance-type statistic between the $p$-degree at the
tail and the $q$-degree at the head of a directed edge, with normalization
\[
\nu_{p,q}=\frac{1}{\sigma_T\sigma_H}
\]
Under degree-preserving rewiring the degree sequence (hence $\sigma_T,\sigma_H$ and the maxima
$d_{\max}^p:=\max_i d_i^p$, $d_{\max}^q:=\max_i d_i^q$) remain fixed. A single swap replaces only
$O(1)$ directed edges, so the assortativity numerator changes by a sum of $O(1)$ degree-products,
each bounded by $d_{\max}^p d_{\max}^q$; after normalization, the one-step variation scale is
therefore
\[
\Lambda_{p,q}\asymp \nu_{p,q}\,d_{\max}^p d_{\max}^q
\]
up to universal constants (and the precise convention used in Def~\ref{def:neumann}). Combining
this leverage estimate with the perturbation-theoretic factor $\kappa_*/\gamma_*$ and the
rewiring budget/participation parameter $r$ (as it enters the operator-norm control of the
cumulative perturbation) gives the multiplicative envelope
\[
M_{p,q}\;\lesssim\; C_{p,q}\,\frac{\kappa_*}{\gamma_*}\,r\,\Lambda_{p,q}
\]
where $C_{p,q}$ is chosen large enough to absorb normalization conventions and any residual slack
terms from the affine comparison.
\enremark
\end{remark}

\section{Local structures}\label{sec:local_structures}
We now turn to local and mesoscopic structures that can pull the leading eigenvector away from the degree proxy. We focus on three canonical patterns: communities (modularity), core-periphery organization, and cycles. For each pattern we specify a degree-preserving rewiring rule that
amplifies the structure, express the resulting change as a cumulative perturbation of the adjacency matrix, and then invoke the Stewart-Sun Perturbation Theorem (Result~\ref{res:stewart-sun}), together with our operator-norm control of the perturbation, to bound the induced degree-eigenvector deviation.

\subsection{Communities}\label{subsec:communities}
\begin{mydef}[Community-contrast statistic]\label{def:community-stat}
Let $\mathbf A\in\{0,1\}^{n\times n}$ be a directed simple graph with
$m:=|\mathcal E(\mathbf A)|$ edges, and fix a partition $\mathcal C=\{C_1,\dots,C_K\}$.
Define the block edge fractions
\[
e_{k\ell}
\;:=\;
\frac{1}{m}\sum_{i\in C_k}\sum_{j\in C_\ell} a_{ij},
\qquad 1\le k,\ell\le K
\]
and the corresponding block marginals
\[
e_{k\cdot}:=\sum_{\ell=1}^K e_{k\ell},
\qquad
e_{\cdot k}:=\sum_{\ell=1}^K e_{\ell k}
\]
The community-contrast of $\mathbf A$ with respect to $\mathcal C$ is
\[
\phi_{\mathrm{com}}(\mathbf A;\mathcal C)
\;:=\;
\sum_{k=1}^K \bigl(e_{kk}-e_{k\cdot}\,e_{\cdot k}\bigr)
\]
This statistic is positive when within-community edge mass exceeds the baseline
predicted by the block marginals, and is small (or negative) when edges are
predominantly between communities.
\enremark
\end{mydef}

\paragraph{Community strengthening rewiring \\}
Let $\mathbf A^{(0)}$ be a neutral baseline (Def~\ref{def:neutral-network}) and fix a
partition of the vertex set into two groups $\mathcal Q_1\cup\mathcal Q_2=\{1,\dots,n\}$, chosen independently of degrees. Impose the angle deviation constraint in Def \ref{def:rewire}. A community-forming swap selects two cross-community edges $(a,b)\in\mathcal E_{12}$ and $(c,d)\in\mathcal E_{21}$ and performs the standard degree-preserving swap
\[
(a,b),\ (c,d)\ \leadsto\ (a,d),\ (c,b),
\qquad a,d\in\mathcal Q_1,\quad b,c\in\mathcal Q_2
\]
thereby converting two cross edges into one within-$\mathcal Q_1$ edge and one within-$\mathcal Q_2$ edge while preserving all in- and out-degrees. 
\\
\indent
Writing $\boldsymbol{\Delta}^{(i)}_{\mathrm{com}}$ for the perturbation matrix of the $i$th such swap and $\boldsymbol{\Omega}_{\mathrm{com}}^{(t)}:=\sum_{i=1}^t\boldsymbol{\Delta}^{(i)}_{\mathrm{com}}$, the rewired adjacency matrix is $\mathbf A^{(t)}=\mathbf A^{(0)}+\boldsymbol{\Omega}_{\mathrm{com}}^{(t)}$. If each vertex participates in at most $r_{\mathrm{com}}$ community-forming swaps, then Proposition~\ref{prop:delta-omega-norm} gives
\[
\bigl\|\boldsymbol{\Omega}_{\mathrm{com}}^{(t)}\bigr\|_2 \;\le\; 2r_{\mathrm{com}}
\]
Hence, whenever $2\kappa r_{\mathrm{com}}/\gamma<1$, Proposition~\ref{prop:deg-evec-SS} along with the neutrality assumption  yields the eigenvector-rotation bound
\[
\sin\theta_{\mathbf d,\mathbf v^{(t)}} \;\le\; \frac{2\kappa r_{\mathrm{com}}}{\gamma}
\]
\paragraph{Extension to many communities\\}
The same construction applies to any partition $\mathcal Q_1,\dots,\mathcal Q_m$ with $m\ge2$: pick any pair of groups $(p,q)$, choose two edges in opposite directions between $\mathcal Q_p$ and $\mathcal Q_q$, and swap them into within-$\mathcal Q_p$ and within-$\mathcal Q_q$ edges. All bounds above remain unchanged, since they depend only on degree preservation and the participation budget $r_{\mathrm{com}}$ through Proposition~\ref{prop:delta-omega-norm} and Proposition~\ref{prop:deg-evec-SS}.

\begin{remark}[Moment bound for the angle via the community statistic]
\label{rem:moments-angle-community}
Let $\phi_{\mathrm{com}}(\mathbf A;\mathcal C)$ be the community-contrast statistic from
Def~\ref{def:community-stat}, computed for a fixed partition
$\mathcal C=\{C_1,\dots,C_K\}$. Along the degree-preserving community-forming rewiring trajectory
$\{\mathbf A^{(t)}\}_{t\ge0}$ define
\[
\phi_t:=\phi_{\mathrm{com}}(\mathbf A^{(t)};\mathcal C),\qquad
\theta_t:=\theta(\mathbf A^{(t)}).
\]
The acceptance rule is $\phi_{\mathrm{com}}$-upper-contour (accepted swaps satisfy
$\phi_{t+1}\ge \phi_t$). If the affine upper comparison assumed in
Proposition~\ref{prop:moment_bounds_angle_from_stat} holds on the reachable set for
$\phi=\phi_{\mathrm{com}}(\cdot;\mathcal C)$, then Corollary~\ref{cor:multiplicative_envelope_abs}
implies the one-moment control
\[
\mathbb E[\theta_t]\;\le\; M_{\mathrm{com}}\Bigl(1+\mathbb E\bigl[|\phi_t|\bigr]\Bigr),
\]
for a uniform constant $M_{\mathrm{com}}$.

\medskip
\noindent\textit{(Community-specific scaling of $M_{\mathrm{com}}$.)}
The partition enters only through its granularity $K$ and block sizes $\{|C_k|\}_{k=1}^K$.
Under degree-preserving swaps the degree sequence is fixed, so the relevant extremal controls are
$d_{\max}^{\mathrm{out}}$ and $d_{\max}^{\mathrm{in}}$. A single community-forming swap replaces
$O(1)$ edges and can change the within-block edge count only by $O(1)$; in weighted form, the
largest mass a swap can redirect into within-community blocks is controlled by the extremal
degree product $d_{\max}^{\mathrm{out}}d_{\max}^{\mathrm{in}}$ (up to the normalization used in
Def~\ref{def:community-stat}). Thus the natural degree-leverage scale is
\[
\Lambda_{\mathrm{com}}\;\asymp\; C_{\mathrm{com}}(\mathcal C)\,d_{\max}^{\mathrm{out}}d_{\max}^{\mathrm{in}},
\]
where $C_{\mathrm{com}}(\mathcal C)$ collects the partition-dependent normalization (e.g.\ factors
depending on $K$ and $\{|C_k|\}$). Combining this leverage estimate with the perturbation factor
$\kappa_*/\gamma_*$ and the community participation/budget parameter $r_{\mathrm{com}}$ gives the
multiplicative envelope
\[
M_{\mathrm{com}}\;\lesssim\; \widetilde C_{\mathrm{com}}(\mathcal C)\,
\frac{\kappa_*}{\gamma_*}\,r_{\mathrm{com}}\,\Lambda_{\mathrm{com}},
\]
with $\widetilde C_{\mathrm{com}}(\mathcal C)$ chosen large enough to absorb normalization
conventions and any residual slack terms from the affine comparison.

\enremark
\end{remark}

\subsection{Core-periphery}\label{subsec:core-periphery}

\begin{mydef}[Core-periphery contrast statistic]\label{def:core-periphery-stat}
Let $\mathbf A\in\{0,1\}^{n\times n}$ be a directed simple graph with
$m:=|\mathcal E(\mathbf A)|$ edges, and fix a core-periphery partition
$\{1,\dots,n\}=\mathcal H\sqcup\mathcal L$. For $X,Y\in\{\mathcal H,\mathcal L\}$ define the block edge
fractions
\[
e_{XY}
\;:=\;
\frac{1}{m}\sum_{i\in X}\sum_{j\in Y} a_{ij}
\]
The core-periphery contrast of $\mathbf A$ (relative to $\mathcal H,\mathcal L$) is
\[
\phi_{\mathrm{cp}}(\mathbf A;\mathcal H,\mathcal L)
\;:=\;
1-2e_{\mathcal L\mathcal L}
\;=\;
\bigl(e_{\mathcal H\mathcal H}+e_{\mathcal H\mathcal L}+e_{\mathcal L\mathcal H}\bigr)
-e_{\mathcal L\mathcal L}
\]
Thus $\phi_{\mathrm{cp}}$ is large when within-periphery density is small, and it decreases as
$\mathcal L\to\mathcal L$ edges accumulate.
\enremark
\end{mydef}

\paragraph{Degree-based partition and a swap\\}

We implement a degree-preserving rewiring that strengthens a core-periphery
pattern: a `core' $\mathcal H$ that is dense internally and well connected to
a `periphery' $\mathcal L$, with relatively few $\mathcal L\to\mathcal L$
links.  Starting from a neutral baseline $\mathbf A$ (Def~\ref{def:neutral-network}),
order vertices by (out-)degree and split them into a high-degree set $\mathcal H$
(core) and a low-degree set $\mathcal L$ (periphery)\footnote{ This degree-biased choice is essential because the hub by definition has more connections than the periphery. The hub must therefore have high-degree nodes as our rewiring process preserves degree.}.  A single core-periphery-forming step is the degree-preserving swap that replaces one periphery-periphery edge with a core-periphery edge: pick
\[
(a\to b)\ \text{with }a,b\in\mathcal L,
\qquad
(c\to d)\ \text{with }c\in\mathcal H
\]
delete these two edges, and add
\[
(a\to d),\qquad (c\to b)
\]
whenever this creates no duplicate edges.  This removes an $\mathcal L\to\mathcal L$ link and forces an $\mathcal H\to\mathcal L$ link, thereby reducing within-periphery density and strengthening core-to-periphery connectivity. Note that $d$ can belong to either $\mathcal L$ or $\mathcal H$, which is precisely what ensures that the density of connections within the hub does not decline as the rewiring process unfolds.

\paragraph{Perturbation and eigenvector control\\}
Let $\boldsymbol{\Delta}^{(t)}_{cp}$ be the perturbation matrix of the
$t^{\mathrm {th}}$ core-periphery swap, and set
\[
\boldsymbol{\Omega}^{(t)}_{\mathrm{cp}}:=\sum_{s=1}^t \boldsymbol{\Delta}^{(s)}_{\mathrm{cp}},
\qquad
\mathbf A^{(t)}=\mathbf A^{(0)}+\boldsymbol{\Omega}^{(t)}_{\mathrm{cp}}
\]
Under an $r_{\mathrm{cp}}$-bounded participation budget, Proposition~\ref{prop:delta-omega-norm}
gives $\|\boldsymbol{\Omega}^{(t)}_{\mathrm{cp}}\|_2\le 2r_{\mathrm{cp}}$. Hence, whenever
$2\kappa r_{\mathrm{cp}}/\gamma<1$, Proposition~\ref{prop:deg-evec-SS} yields
\[
\sin\bigl(\theta_{\mathbf d,\mathbf v^{(t)}}\bigr) \ \le\ \frac{2\kappa r_{\mathrm{cp}}}{\gamma}
\]
Since $\mathbf A^{(t)}$ shares the same in- and out-degree sequences as the neutral baseline,
Proposition~\ref{prop:neutral-degree-evec} then transfers this control to the
degree--eigenvector misalignment: for modest $r_{\mathrm{cp}}$ (relative to $\gamma/\kappa$),
degree remains a reliable proxy even as the rewiring amplifies a core-periphery pattern.

\paragraph{Fractal core-periphery generalization\\}
The same degree-preserving swap can be iterated on nested partitions to
create a hierarchical core-periphery pattern. Start with the degree-based split
$\{1,\dots,n\}=\mathcal H^{(0)}\cup\mathcal L^{(0)}$. Apply the core-periphery-forming rewiring
to strengthen connectivity within $\mathcal H^{(0)}$ and from $\mathcal H^{(0)}$
to $\mathcal L^{(0)}$, producing $\mathbf A^{(t)}=\mathbf A^{(0)}+\boldsymbol{\Omega}^{(0)}$
after the level-$0$ swaps.

To generate additional levels, recursively refine the periphery: at level
$\ell\ge 1$, partition each periphery block from level $\ell-1$ into sub-blocks,
split each sub-block into a local high-degree set (local core) and low-degree
set (local periphery), and apply the same degree-preserving core-periphery
swap restricted to edges whose endpoints lie inside that sub-block. Let
$r_{\mathrm{cp}}^{(\ell)}$ be the per-node participation budget at level $\ell$,
and let $\boldsymbol{\Omega}_{\mathrm{cp}}^{(\ell)}$ be the cumulative
perturbation contributed by all swaps at that level.

Since every swap is a degree-preserving swap, the operator-norm control is
identical at every level, by Proposition~\ref{prop:delta-omega-norm},
\[
\bigl\|\boldsymbol{\Omega}_{\mathrm{cp}}^{(\ell)}\bigr\|_2 \;\le\; 2\,r_{\mathrm{cp}}^{(\ell)}
\]
After $L$ levels, the total perturbation is
$\boldsymbol{\Omega}^{\mathrm{(tot)}}_{\mathrm{cp}}:=\sum_{\ell=0}^{L}\boldsymbol{\Omega}^{(\ell)}_{\mathrm{cp}}$,
and each node participates in at most
$r_{\mathrm{tot}}:=\sum_{\ell=0}^{L} r_{\mathrm{cp}}^{(\ell)}$ swaps overall, hence again
(Proposition~\ref{prop:delta-omega-norm})
\[
\bigl\|\boldsymbol{\Omega}^{\mathrm{(tot)}}_{\mathrm{cp}}\bigr\|_2 \;\le\; 2\,r^{\mathrm{tot}}_{cp}
\]
Therefore, whenever $2\kappa r_{\mathrm{tot}}/\gamma<1$,
Proposition~\ref{prop:deg-evec-SS} yields the single-line eigenvector-rotation bound
\[
\sin\theta^{\mathrm{(tot)}}_{\mathrm{cp}}
\;\le\;
\frac{2\kappa\,r^{\mathrm{tot}}_{\mathrm{cp}}}{\gamma}
\]
Because every swap preserves the in- and out-degree sequences, the degree proxy
$\mathbf d$ is unchanged throughout, and in the neutral regime
(Proposition~\ref{prop:neutral-degree-evec}) this implies that even a multi-level
core-periphery construction keeps the leading eigenvector close to degree as
long as the total per-node swap budget $r_{\mathrm{tot}}$ remains modest
relative to $\gamma/\kappa$.

\begin{remark}[Moment bound for the angle via core--periphery contrast]
\label{rem:moments-angle-coreperiph}
Let $\phi_{\mathrm{cp}}(\mathbf A;\mathcal H,\mathcal L)$ be the core--periphery contrast from
Def~\ref{def:core-periphery-stat}, computed using the degree-based partition
$(\mathcal H,\mathcal L)$. Along the degree-preserving core--periphery rewiring trajectory
$\{\mathbf A^{(t)}\}_{t\ge0}$ define
\[
\phi_t:=\phi_{\mathrm{cp}}(\mathbf A^{(t)};\mathcal H,\mathcal L),\qquad
\theta_t:=\theta(\mathbf A^{(t)})
\]
By construction, each accepted swap deletes an $\mathcal L\!\to\!\mathcal L$ edge and does not
create a new $\mathcal L\!\to\!\mathcal L$ edge; thus $e_{\mathcal L\mathcal L}$ weakly
decreases, and since $\phi_{\mathrm{cp}}(\cdot;\mathcal H,\mathcal L)$ is monotone in
$e_{\mathcal L\mathcal L}$ when degrees are fixed, we have $\phi_{t+1}\ge \phi_t$ pathwise.
If the affine upper comparison assumed in Proposition~\ref{prop:moment_bounds_angle_from_stat}
holds on the reachable set for $\phi=\phi_{\mathrm{cp}}(\cdot;\mathcal H,\mathcal L)$, then
Corollary~\ref{cor:multiplicative_envelope_abs} implies the one-moment control
\[
\mathbb E[\theta_t]\;\le\; M_{\mathrm{cp}}\Bigl(1+\mathbb E\bigl[|\phi_t|\bigr]\Bigr)
\]
for a uniform constant $M_{\mathrm{cp}}$.

\medskip
\noindent\textit{(Core--periphery-specific scaling of $M_{\mathrm{cp}}$.)}
Here the relevant budget is $r_{\mathrm{cp}}$ (as it enters the operator-norm control of the
cumulative perturbation), and the partition enters only through $|\mathcal H|,|\mathcal L|$.
A single accepted swap changes only $O(1)$ directed edges; in terms of the contrast statistic,
the largest change comes from redirecting mass away from $\mathcal L\!\to\!\mathcal L$ and into
blocks involving $\mathcal H$, which is controlled by the extremal degree product
$d_{\max}^{\mathrm{out}}d_{\max}^{\mathrm{in}}$ (up to the normalization in
Def~\ref{def:core-periphery-stat}). Thus the natural leverage scale is
\[
\Lambda_{\mathrm{cp}}\;\asymp\; C_{\mathrm{cp}}(\mathcal H,\mathcal L)\,
d_{\max}^{\mathrm{out}}d_{\max}^{\mathrm{in}}
\]
with $C_{\mathrm{cp}}(\mathcal H,\mathcal L)$ collecting the block-size/normalization factors. Combining this with the perturbation factor $\kappa_*/\gamma_*$ and the budget $r_{\mathrm{cp}}$ gives the multiplicative envelope
\[
M_{\mathrm{cp}}\;\lesssim\; \widetilde C_{\mathrm{cp}}(\mathcal H,\mathcal L)\,
\frac{\kappa_*}{\gamma_*}\,r_{\mathrm{cp}}\,\Lambda_{\mathrm{cp}}
\]
where $\widetilde C_{\mathrm{cp}}(\mathcal H,\mathcal L)$ is chosen large enough to absorb
normalization conventions and any residual slack terms from the affine comparison.

\enremark
\end{remark}

\subsection{Cycles}\label{subsec:cycles}

\begin{mydef}[Cycle-density statistic ($k$-cycle participation)]
\label{def:cycle_stat}
A directed $k$-cycle is a simple motif
$i_1\!\to i_2\!\to\cdots\to i_k\!\to i_1$ with distinct vertices. To measure a cycle,  fix an integer $k\ge 3$ and let $\mathbf A\in\{0,1\}^{n\times n}$ be the adjacency
matrix of a directed simple graph on vertex set $\{1,\dots,n\}$.
A directed $k$-cycle is an ordered $k$-tuple of distinct vertices
$(i_1,\dots,i_k)$ such that
\[
(i_1\!\to i_2),\ (i_2\!\to i_3),\ \dots,\ (i_{k-1}\!\to i_k),\ (i_k\!\to i_1)\in \mathcal E(\mathbf A)
\]
Let $C_k(\mathbf A)$ denote the total number of directed $k$-cycles in $\mathbf A$
(counted up to cyclic rotation).\footnote{Equivalently, $C_k(\mathbf A)=\frac{1}{k}\,
\#\{(i_1,\dots,i_k):\ (i_1,\dots,i_k)\ \text{forms a directed $k$-cycle}\}$.
Any consistent counting convention (e.g.\ counting ordered cycles and dividing by $k$) may be used.}
Define the $k$-cycle density by
\[
\phi_{k\text{-}\mathrm{cyc}}(\mathbf A)
\;:=\;
\frac{C_k(\mathbf A)}{(n)_k/k}
\;\in\;[0,1],
\qquad (n)_k:=n(n-1)\cdots(n-k+1)
\]
Here $(n)_k/k$ is the number of distinct directed $k$-cycles in the complete
directed graph (with no self-loops), so $\phi_{k\text{-}\mathrm{cyc}}(\mathbf A)$
is the fraction of all possible directed $k$-cycles that are present in $\mathbf A$.
\enremark
\end{mydef}

 We now describe degree-preserving swaps that create short
cycles (triangles and, more generally, $k$-cycles) from the neutral benchmark,
and then invoke the same perturbation machinery as before: each
cycle-forming move is a degree-preserving swap, hence contributes a rank-one
$\{-1,0,1\}$ perturbation with spectral norm at most $2$. Naturally then the cumulative
perturbation will be controlled by Proposition~\ref{prop:delta-omega-norm} and the
resulting eigenvector rotation by the Stewart--Sun bound
(Result~\ref{res:stewart-sun}).

\paragraph{Triangle-forming swap \\}

Given $\mathbf A$, choose distinct $a,b,c$ with $(a\!\to b),(b\!\to c)\in\mathcal E(\mathbf A)$
and $(c\!\to a)\notin\mathcal E(\mathbf A)$. Pick edges $(c\!\to d)$ and $(e\!\to a)$ with
$d,e\notin\{a,b,c\}$ such that $(c\!\to a)$ and $(e\!\to d)$ are admissible (no self-loops,
not already present), and perform the swap
\[
(c\!\to d,\ e\!\to a)\quad\leadsto\quad(c\!\to a,\ e\!\to d)
\]
which preserves in- and out-degrees and closes the triangle
$a\!\to b\!\to c\!\to a$. Writing $\boldsymbol{\Omega}^{(t)}_{\mathrm{tri}}$ for the cumulative
perturbation after $t$ triangle-forming swaps and assuming each node participates in at
most $r_{\mathrm{tri}}$ such swaps, Proposition~\ref{prop:delta-omega-norm} gives
\[
\bigl\|\boldsymbol{\Omega}^{(t)}_{\mathrm{tri}}\bigr\|_2\ \le\ 2r_{\mathrm{tri}}
\]
Therefore, whenever $2r_{\mathrm{tri}}<\gamma/\kappa$, Result~\ref{res:stewart-sun} and
Proposition~\ref{prop:deg-evec-SS} yield the usual proxy-stability bound
\[
\sin\angle\!\bigl(\mathbf d,\mathbf v^{(t)}\bigr)\ \lesssim\ \frac{2\kappa r_{\mathrm{tri}}}{\gamma}
\]
where $\mathbf v^{(t)}$ is the unit leading eigenvector of $\mathbf A^{(t)}$ and $\mathbf d$ is
the (unit) degree proxy fixed by the degree sequence.

\paragraph{$\mathbf{k}$-cycles \\}
Fix $k\ge3$. To close an almost $k$-cycle
$i_1\!\to\cdots\to i_k$ with missing edge $(i_k\!\to i_1)$, perform the analogous swap that
adds $(i_k\!\to i_1)$ while deleting one outgoing edge of $i_k$ and one incoming edge of $i_1$
(and reconnecting their other endpoints). Each $k$-cycle-forming move is again a single
degree-preserving swap, hence the same bounds hold with $r_{\mathrm{tri}}$ replaced by the
per-node participation budget $r_{k\text{-}\mathrm{cyc}}$:
\[
\bigl\|\boldsymbol{\Omega}^{(t)}_{k\text{-}\mathrm{cyc}}\bigr\|_2\ \le\ 2r_{k\text{-}\mathrm{cyc}},
\qquad
\sin\angle\!\bigl(\mathbf d,\mathbf v^{(t)}\bigr)\ \lesssim\ \frac{2\kappa r_{k\text{-}\mathrm{cyc}}}{\gamma}
\]
whenever $2r_{k\text{-}\mathrm{cyc}}<\gamma/\kappa$.

\paragraph{Growing longer cycles\\}
When `almost' $k$-cycles are scarce, one can instead grow a triangle into a longer cycle by a
constant number of additional swaps per unit increase in length (rerouting one cycle edge
through a fresh directed length-$2$ chain disjoint from existing cycles). If creating a single
length-$k$ cycle from a triangle uses at most $c_k=O(k)$ additional swapes, and each node
participates in at most $r_k$ such growth steps, then the total perturbation satisfies the crude
accumulation bound
\[
\bigl\|\boldsymbol{\Omega}^{\mathrm{tot}}_{k}\bigr\|_2
\ \lesssim\ 2r_{\mathrm{tri}} + 2c_k r_k
\]
and hence the Stewart--Sun angle control scales accordingly:
\[
\sin\angle\!\bigl(\mathbf d,\mathbf v^{(k)}\bigr)
\ \lesssim\ \frac{2\kappa}{\gamma}\,\bigl(r_{\mathrm{tri}}+c_k r_k\bigr)
\]
so longer cycles are `costlier' because the number of required local swaps grows with $k$.

\begin{remark}[Moment bound for the angle via the $k$-cycle statistic]
\label{rem:moments-angle-cycles}
Fix $k\ge 3$ and let $\phi_{k\text{-}\mathrm{cyc}}(\mathbf A)$ be the cycle statistic from
Def~\ref{def:cycle_stat}. Along the $k$-cycle--forming, degree-preserving rewiring trajectory
$\{\mathbf A^{(t)}\}_{t\ge0}$ define
\[
\phi_t:=\phi_{k\text{-}\mathrm{cyc}}(\mathbf A^{(t)}),\qquad
\theta_t:=\theta(\mathbf A^{(t)})
\]
The acceptance rule is $\phi_{k\text{-}\mathrm{cyc}}$-upper-contour (accepted swaps satisfy
$\phi_{t+1}\ge \phi_t$), and by construction $0\le \phi_t\le 1$. If the affine upper comparison
assumed in Proposition~\ref{prop:moment_bounds_angle_from_stat} holds on the reachable set for
$\phi=\phi_{k\text{-}\mathrm{cyc}}$, then Corollary~\ref{cor:multiplicative_envelope_abs} yields
\[
\mathbb E[\theta_t]\;\le\; M_{k\text{-}\mathrm{cyc}}\Bigl(1+\mathbb E\bigl[|\phi_t|\bigr]\Bigr)
\;=\; M_{k\text{-}\mathrm{cyc}}\Bigl(1+\mathbb E[\phi_t]\Bigr)
\]
for a uniform constant $M_{k\text{-}\mathrm{cyc}}$.

\medskip
\noindent\textit{($k$-cycle-specific scaling of $M_{k\text{-}\mathrm{cyc}}$.)}
The statistic depends only on the local closure of length-$k$ motifs, so a single accepted swap
affects the cycle count through $O_k(1)$ edge incidences (a combinatorial constant depending only
on $k$). Under degree preservation the extremal local edge-mass available for closing cycles is
controlled by $d_{\max}^{\mathrm{out}}$ and $d_{\max}^{\mathrm{in}}$; in worst case, closing
short cycles concentrates around high-degree vertices, giving the leverage scale
\[
\Lambda_{k\text{-}\mathrm{cyc}}\;\asymp\; C_k\,d_{\max}^{\mathrm{out}}d_{\max}^{\mathrm{in}},
\]
where $C_k$ absorbs the $k$-dependent combinatorial factor and the normalization convention used
in $\phi_{k\text{-}\mathrm{cyc}}$. Combining this with the perturbation factor $\kappa_*/\gamma_*$
and the $k$-cycle participation/budget parameter $r_{k\text{-}\mathrm{cyc}}$ gives the
multiplicative envelope
\[
M_{k\text{-}\mathrm{cyc}}\;\lesssim\; \widetilde C_k\,\frac{\kappa_*}{\gamma_*}\,
r_{k\text{-}\mathrm{cyc}}\,\Lambda_{k\text{-}\mathrm{cyc}}
\]
with $\widetilde C_k$ chosen large enough to absorb normalization and any residual slack from the
affine comparison.

\enremark
\end{remark}

\newpage
\section{Power-law tails and statistic-specific bounds}
\label{sec:bound_diff}

Many large real-world networks exhibit pronounced degree heterogeneity, often well approximated in
their upper tails by power laws. For our purposes, the importance of this empirical regularity is
not merely descriptive. It identifies the regime in which a small number of vertices carry a
disproportionate share of edge-ends, and hence the regime in which degree-preserving rewiring can
have especially uneven and high-leverage effects on global network statistics. This is also the
regime in which the constants appearing in our moment bounds become most informative.

A practically important point is that these constants are not universal. All four statistics we
study share the same spectral backbone: eigenvector responsiveness enters through the conditioning
and separation parameters $(\kappa,\gamma)$, while the cumulative size of the rewiring enters
through the bounded-participation operator-norm budget $r$. Where the statistics differ is in how a
change in the chosen global statistic $\phi$ translates into a bound on the deviation angle
$\theta(\mathbf A)$. The ratio $\kappa/\gamma$ captures the common spectral part of this relation
by measuring how fragile the Perron eigenvector is to perturbations of a given operator norm. What
varies across statistics is the combinatorial part: how much such perturbation, and hence how many
degree-preserving edits under the $r$-bounded participation rule, are needed to move the statistic
by a given amount. This difference is reflected in the affine constants appearing in the moment
bounds. Their formal role is the same across statistics, but their magnitude is not, since it
depends on how sensitive the statistic is to degree-preserving swaps and on how much eigenvector
rotation those swaps can induce. In the power-law regime, this statistic-specific component becomes
especially transparent, because the dominant contribution comes from extremal degrees. If
\[
d_{\max}^{\mathrm{out}}\asymp d_{\max}^{\mathrm{in}}\asymp d_{\max}\asymp n^{1/\alpha},
\qquad \alpha>1,
\]
up to slowly varying factors, then each statistic acquires a common heavy-tail scaling of order
$d_{\max}^2\asymp n^{2/\alpha}$, with the remaining constant determined by the particular feature
being rewired. We now examine how this works for assortativity, community structure,
core--periphery structure, and cycles.

\subsection*{Assortativity}
For assortativity, the relevant constants are governed by endpoint-degree dispersion. Newman
assortativity is a normalized covariance across edges, and $\phi_{p,q}$ correlates the $p$-degree
at the tail of an edge with the $q$-degree at its head. A degree-preserving swap cannot alter the
endpoint-degree multisets; it can only change which tails are paired with which heads. That is the
only margin along which assortativity moves. The normalization
\[
\nu_{p,q}=\frac{1}{\sigma_T\sigma_H}
\]
rescales the raw degree-product covariance into a dimensionless correlation, where $\sigma_T$ and
$\sigma_H$ are the standard deviations of the tail- and head-side endpoint-degree samples. Along
the rewiring trajectory these dispersions, and hence $\nu_{p,q}$, remain fixed, so rewiring acts
only through changes in cross-edge covariance. Because each swap re-pairs only a constant number of
endpoints, the per-swap movement in $\phi_{p,q}$ is driven by the upper tail. Most swaps touch
only moderate-degree vertices and have little effect, but a swap involving a rare high-degree node
can generate a much larger jump in the degree-product sum. In the power-law regime one has
\[
\Lambda_{p,q}\asymp \nu_{p,q} d_{\max}^2
\]
and hence
\[
M_{p,q}\;\lesssim\; \widetilde C_{p,q}\,\nu_{p,q}\,
\frac{\kappa_*}{\gamma_*}\,r\,n^{2/\alpha}.
\]
Thus smaller $\alpha$, corresponding to heavier tails, enlarges the time-uniform envelope for
$\mathbb E[\theta_t]$ by increasing the leverage of extremal endpoint degrees.

\subsection*{Communities}
For community structure, the constants are shaped by the chosen partition because
$\phi_{\mathrm{com}}$ records how much edge mass lies within blocks rather than across them. A
degree-preserving swap cannot change how many stubs each vertex contributes, so the only way to
increase $\phi_{\mathrm{com}}$ is to reroute existing cross-block mass into within-block locations
while keeping all row and column sums fixed. How fast this can happen depends on two ingredients.
First, the partition geometry matters. Smaller or more numerous communities provide less
within-block capacity, so the process saturates earlier. Larger or more balanced blocks provide
more admissible within-block destinations. Second, degree heterogeneity matters because most of the
movable mass is carried by a relatively small set of high-degree vertices. If these hubs have many
admissible within-block partners, cross-block mass can be converted quickly; if not, gains in
$\phi_{\mathrm{com}}$ stall even when many swaps remain available. The affine constants therefore
depend jointly on partition geometry and the degree tail: the partition determines the available
within-block capacity, while the tail determines how concentrated the movable mass is. In the
power-law regime,
\[
\Lambda_{\mathrm{com}}
\asymp
C_{\mathrm{com}}(\mathcal C)\,d_{\max}^2
\]
and therefore
\[
M_{\mathrm{com}}
\;\lesssim\;
\widehat C_{\mathrm{com}}(\mathcal C)\,
\frac{\kappa_*}{\gamma_*}\,
r_{\mathrm{com}}\,
n^{2/\alpha}.
\]
Hence heavier tails enlarge the time-uniform envelope for $\mathbb E[\theta_t]$ through greater
extremal-degree leverage under degree-preserving rewiring.

\subsection*{Core--periphery}
Core--periphery has a similar mass-transfer logic, but the target is asymmetric. The statistic
$\phi_{\mathrm{cp}}$ rises primarily by draining mass from the $\mathcal L\!\to\!\mathcal L$ block
and redirecting it so that periphery endpoints connect through the core $\mathcal H$. Since
vertexwise in- and out-degrees are fixed, rewiring cannot change how much peripheral stub mass
exists; it can only change where that mass lands. The relevant constants are therefore governed by
the split $(\mathcal H,\mathcal L)$ and by the absorptive capacity of the core. If the core has
enough degree mass, periphery--periphery edges can be progressively reattached through
$\mathcal H$; if not, some $\mathcal L\!\to\!\mathcal L$ residue is unavoidable. Degree
dispersion matters because in heavy-tailed networks a few very high-degree core vertices act as
large-capacity hubs, allowing more peripheral mass to be redirected before saturation occurs. This
gives core--periphery relatively high leverage: a modest number of well-placed swaps can rapidly
concentrate paths through the core, to which the Perron vector is especially responsive. In the
power-law regime,
\[
\Lambda_{\mathrm{cp}}
\asymp
C_{\mathrm{cp}}(\mathcal H,\mathcal L)\,d_{\max}^2
\]
and hence
\[
M_{\mathrm{cp}}
\;\lesssim\;
\widehat C_{\mathrm{cp}}(\mathcal H,\mathcal L)\,
\frac{\kappa_*}{\gamma_*}\,
r_{\mathrm{cp}}\,
n^{2/\alpha}.
\]
So here too heavier tails enlarge the time-uniform envelope for $\mathbb E[\theta_t]$ through
extremal-degree leverage under degree-preserving rewiring.

\subsection*{Cycles}
For $k$-cycles, $\phi_{k\text{-}\mathrm{cyc}}$ is motif-based, so the relevant rewiring is local. A
single accepted swap can create a new short cycle, or complete an incipient one, by modifying only
a constant number of edges. This locality makes cycles a useful stress test for eigenvector
robustness: a small motif inserted in a small region may induce noticeable localized eigenvector
distortion even when coarse global summaries move slowly. Degree heterogeneity again matters
combinatorially. High-degree vertices provide many potential neighbors, so swaps involving hubs
often make it easier to close cycles because there are more admissible ways to complete the
required adjacency pattern under degree preservation. Longer cycles are usually costlier because
building a $k$-cycle requires coordinating adjacency among $O(k)$ vertices and often multiple
accepted swaps. But heavy tails mitigate this cost by supplying hubs as attachment points, making
long-cycle assembly far easier than in thin-tailed graphs. Accordingly, in the power-law regime,
\[
\Lambda_{k\text{-}\mathrm{cyc}}
\asymp
C_k d_{\max}^2
\]
and therefore
\[
M_{k\text{-}\mathrm{cyc}}
\;\lesssim\;
\widehat C_k\,
\frac{\kappa_*}{\gamma_*}\,
r_{k\text{-}\mathrm{cyc}}\,
n^{2/\alpha}.
\]
Thus heavier tails enlarge the time-uniform envelope for $\mathbb E[\theta_t]$ by increasing the
combinatorial leverage available for closing cycles around hubs.

Across all four cases, the same mechanism is at work. Heavy degree tails determine how much
structural change can be generated with a fixed degree-preserving perturbation budget because a
small set of hubs carries a disproportionate share of edge-ends. This makes the evolution uneven:
swaps involving hubs can move the statistic substantially, whereas most swaps that avoid them do
relatively little. The displayed bounds capture this through the common scaling
\[
M\;\lesssim\;(\text{statistic-specific constant})
\cdot
\frac{\kappa_*}{\gamma_*}\cdot r\cdot n^{2/\alpha},
\]
where the statistic-specific constant reflects whether one is re-pairing endpoints, redirecting
mass across blocks, draining periphery--periphery edges through the core, or closing local motifs.
The factor $n^{2/\alpha}$ is the signature of extremal-degree leverage.

There is, however, a second channel through which heavy tails can matter. They may also affect the
spectral sensitivity parameters themselves. Hub-centered rewiring tends to build several forms of
structure around the same small set of vertices, which can worsen eigenbasis conditioning or reduce
modal separation. In the language of the Stewart--Sun bound, this means that heavy-tailed rewiring
may increase $\kappa$ and/or reduce $\gamma$, amplifying the eigenvector response to a given
perturbation.\footnote{In general $\kappa(A)$ and $\gamma(A)$ are not invariant under
degree-preserving rewiring: they depend on the current adjacency matrix and may drift along the
trajectory. Accordingly, as in earlier remarks, we interpret the bound uniformly on a reachable
region $S_0$, working with the time-invariant descriptors
\[
\kappa_*:=\sup_{A\in S_0}\kappa(A),
\qquad
\gamma_*:=\inf_{A\in S_0}\gamma(A)>0.
\]
In heavy-tailed settings these worst-case descriptors can deteriorate, since hub-centered local
structure can increase modal interaction and weaken eigenvalue separation.} Heavy tails therefore
operate through two distinct channels. The direct channel is combinatorial: they increase the
leverage with which degree-preserving swaps can move the chosen statistic. The indirect channel is
spectral: they may also make the leading eigenvector more fragile by worsening conditioning or
shrinking separation. The net effect on the angle--moment bounds is therefore not mechanically
monotone; it depends on which channel dominates in the regime under study.

\section{Concluding remarks} \label{sec:concluding}
 Applied researchers working in data-scarce environments seldom have the information necessary to compute the eigenvectors of the network. Since the eigenvector is a good description of the systemic significance of nodes---whether it be  firms about to foreclose or humans about to transmit a disease---the researcher is left with having to proxy it. Which he typically does with the degree vector\footnote{ Typically, the number of connections of a node is more easily known than the entire network structure.}. Over the years, a certain `folk theorem' has emerged around this substitution, which claims that the procedure is not all too bad when the network has neutral degree-mixing and lacks meso structures.  This folk theorem, however, does not tell us anything about the error introduced by the substitution in networks that violate the neutrality assumptions. This paper presents analytical bounds on the error that is born from substituting the degree-vector for the eigenvector in assortative networks with a modicum of meso structures. We start from a neutral benchmark in which degree and eigenvector centrality align, and then introduce four common departures from neutrality: degree assortativity, community structure, core-periphery organization, and short directed cycles. For each departure, we describe a degree-preserving rewiring mechanism that strengthens the corresponding structure, construct the induced perturbation of the adjacency matrix, and translate it into an upper bound on the deviation angle between the degree vector and the leading eigenvector via the Stewart-Sun Perturbation Theorem. We broaden this methodology to encompass weighted networks as well, showing that continuous strength-preserving weight transfers yield analogous analytical bounds. Furthermore, our theoretical framework is complemented by numerical simulations that illustrate the real-time dynamics of eigenvector rotation, explicitly mapping the wandering and targeted drifting of the deviation angle under various structural perturbations. With this procedure, we placed a boundary around an error term of some empirical importance in economics, epidemiology, opinion dynamics, and other applied sciences.
\\
\indent
Note that our analytical results are only reasonable upper bounds. In principle, local structures can be introduced without disturbing the alignment between degree and eigenvector.  In fact, one could even place local structures to bring the two closer.  The reason is simple: insofar as the preponderance of local structure accrues to (or is concentrated around) high-degree nodes, its presence need not pull the Perron eigenvector away from the degree proxy.  One reason for such an aligned preponderance is the combinatorial fact that higher-degree nodes have a greater possibility of participating in a wide variety of local structures. In fact, some local structures---such as long cycles---are difficult to construct without the participation of high-degree nodes.  So at one extreme, local structures are no trouble at all. At the other extreme, they can wholly misalign the degree and the eigenvector. To study this misalignment, we ensure that the local structures created through our rewiring process maximally disturb the alignment between degree and eigenvector. We do this through two assumptions. The first of which is that no node can participate in more than $r$ rewiring with $r \ll n$\footnote{This assumption is required for the Stewart-Sun perturbation bound to hold.}.   Naturally, this means that high-degree nodes cannot participate in as many local structures as they might have otherwise. The limit curtails the number of local structures that can form around high-degree nodes, thereby causing greater deviation of the eigenvector from the degree vector. The bound on the angle of deviation derived using the Stewart-Sun procedure is therefore naturally an envelope, particularly in the case of heavy-tailed networks where the limit $r$ will almost certainly be binding. The second assumption involves putting structure on the evolution of the angle of deviation in response to the evolution of statistics that measure assortativity and local structures. We assume that the expectation of the angle of deviation increases with the concerned network statistics. This structural assumption allows bound the moments of the angle of deviation with the moments of the concerned network statistics. Naturally, this expectational assumption means that this moment-bound, too, is an envelope.  All of this is to say that our procedure establishes a maximal angle of deviation between the degree vector and the eigenvector, a `reasonable' worst-case scenario if you will\footnote{We say `reasonable' worst-case because one could develop even `worse' worse-case scenarios by implanting local structures at particular chosen locations in a graph.}. In real-world networks, the error induced by substituting the degree vector with the eigenvector could be markedly smaller.

\newpage
\appendix
\section{Extension to Weighted Directed Networks}
\label{app:weighted_networks}
The analysis in the main text is developed for unweighted directed graphs, but many real-world
networks of interest are naturally weighted, so it is important to extend the results to that
setting. The difficulty is that our discrete degree-preserving edge-swapping procedure does not
carry over literally. In a weighted graph, if one insists on swapping whole edges, it becomes
combinatorially implausible to find pairs of links with exactly matching weights. We resolve this
difficulty by reinterpreting a discrete edge as a bundle of finer weighted units and replacing
literal edge swaps by transfers of weight. Once the problem is
framed in this way, there is always scope to reallocate a sufficiently small amount of mass so as
to move the chosen statistic in the desired direction, while keeping in- and out-strengths fixed. 

Let $\mathbf A\in\mathbb R_{\ge 0}^{n\times n}$ be a weighted adjacency matrix. The degree
sequences are replaced by the in- and out-strength sequences
\[
s_i^{\mathrm{in}}:=\sum_j a_{ji},
\qquad
s_i^{\mathrm{out}}:=\sum_j a_{ij}
\]
and the degree proxy vector $\mathbf d$ by the corresponding $\ell_2$-normalized strength vector
$\mathbf s$. Now fix a weighted matrix $\mathbf A^{(t)}$, choose two directed edges $(a\to b)$ and
$(c\to d)$ with positive weights $a_{ab},a_{cd}>0$, and let
\[
0<\delta\le \min\{a_{ab},a_{cd}\}
\]
A strength-preserving transfer of size $\delta$ subtracts $\delta$ from the existing edges and adds
$\delta$ to the cross-edges:
\[
a_{ab}\mapsto a_{ab}-\delta,\qquad a_{cd}\mapsto a_{cd}-\delta
\]
\[
a_{ad}\mapsto a_{ad}+\delta,\qquad a_{cb}\mapsto a_{cb}+\delta
\]
This is the weighted analogue of the degree-preserving swap, but it is better understood as the
elementary move of a continuous rewiring dynamics on the set of nonnegative matrices with fixed
in- and out-strength sequences. The associated perturbation matrix $\boldsymbol\Delta^{(t)}$ has
exactly four nonzero entries, equal to $\pm\delta$, and its row and column sums are zero. Hence
both the out-strength and in-strength sequences remain invariant under every transfer.

Suppose that across the first $t$ transfers no node reallocates more than a total
weight budget of $D$. That is, if node $u$ participates in transfers indexed by $s$, then
\[
\sum_s \delta^{(s)}\le D
\]
Let
\[
\boldsymbol\Omega^{(t)}:=\sum_{s=1}^t \boldsymbol\Delta^{(s)}
\]
denote the cumulative perturbation. Each time node $u$ participates in a transfer of size
$\delta^{(s)}$, it contributes at most two entries of magnitude $\delta^{(s)}$ to row $u$ and at
most two such entries to column $u$. Therefore
\[
\max_u \sum_j \bigl|\Omega_{uj}^{(t)}\bigr|\le 2D,
\qquad
\max_u \sum_i \bigl|\Omega_{iu}^{(t)}\bigr|\le 2D
\]
so that
\[
\|\boldsymbol\Omega^{(t)}\|_\infty\le 2D,
\qquad
\|\boldsymbol\Omega^{(t)}\|_1\le 2D
\]
By the standard inequality between induced norms,
\[
\|\boldsymbol\Omega^{(t)}\|_2
\le
\sqrt{\|\boldsymbol\Omega^{(t)}\|_1\|\boldsymbol\Omega^{(t)}\|_\infty}
\le 2D
\]
Substituting this bound into the Stewart--Sun inequality yields
\[
\sin\theta_{\mathbf s,\mathbf v^{(t)}}\le \frac{2\kappa D}{\gamma}
\]
for either the left or right Perron eigenvector $\mathbf v^{(t)}$, whenever
\[
2\kappa D<\gamma
\]

The weighted setting therefore preserves the perturbation logic of the unweighted one. Uniform control of eigenvector rotation still
comes from bounded cumulative participation, but participation is now measured by total mass
reallocated rather than by the number of discrete edits. \footnote{To apply the
moment-transfer logic of Proposition~\ref{prop:moment_bounds_angle_from_stat}, one defines weighted
counterparts of the earlier unweighted statistics by replacing edge counts with edge weights and
degrees with strengths. Thus weighted assortativity is based on weighted endpoint-strength
pairings, weighted community and core--periphery statistics on the corresponding block shares of
total weight, and weighted cycle statistics on weighted motif counts such as $\mathrm{tr}(\mathbf
A^k)$. In each case, the statistic varies continuously with the mass reallocated, so the extension
from the unweighted setting goes through without hindrance.}

\newpage
\section{Numerical Simulations}
\label{app:simulations}

All our analytical results take the form of bounds: they show how the angle of deviation between
degree and the Perron eigenvector is controlled by the values of assortativity, community,
core--periphery, and cycle statistics. In some empirical applications, however, one may wish to go
beyond such bounds and obtain a sense of the realized deviation angle associated with a given value
of one of these statistics. The purpose of this appendix is to study the
angle--statistic relationship directly by simulation. Because the proofs in the paper are
constructive, these experiments are straightforward to implement, i.e. the simulation procedures are the
same statistic-driven, degree-preserving rewiring algorithms used in the proofs. Throughout, we
report results from $100$ independent simulations for each statistic. Unless otherwise stated, all
runs begin from unweighted directed Erdos--Renyi graphs with $n=10{,}000$ vertices and target mean
out-degree $\bar d\approx 12$, implemented by $p=\bar d/(n-1)$. In what follows, we
report the degree--eigenvector deviation angles in degrees:
\[
\theta_R(\mathrm{out})
:=
\frac{180}{\pi}\,\theta_{\mathbf v^{\mathrm R}(\mathbf A),\,\mathbf d^{\mathrm{out}}}
\qquad\text{and}\qquad
\theta_L(\mathrm{in})
:=
\frac{180}{\pi}\,\theta_{\mathbf v^{\mathrm L}(\mathbf A),\,\mathbf d^{\mathrm{in}}}
\]

\subsection{Assortativity rewiring}

\begin{figure}[H]
\centering
\scalebox{0.8}{\includegraphics[width=\textwidth]{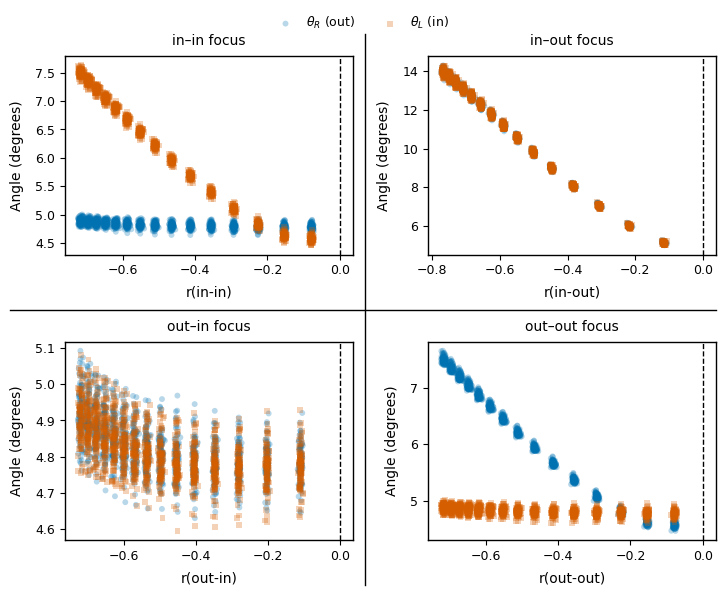}}
\caption{Assortativity experiments : 
Each panel fixes a focus (in--in, in--out, out--in, out--out). 
The x-axis is the corresponding 
degree-endpoint correlation $r(\cdot)$ computed across directed edges; the dashed vertical line marks $r=0$. 
The y-axis shows the deviation angles $\theta_R(\mathrm{out})$ (right eigenvector vs out-degree; blue circles) 
and $\theta_L(\mathrm{in})$ (left eigenvector vs in-degree; orange squares), in degrees.} %
\label{fig:app-assortativity}
\end{figure}

Each point in Figure~\ref{fig:app-assortativity} is a logged snapshot along a degree-preserving
rewiring trajectory: the horizontal axis records a chosen assortativity coefficient $r(\cdot)$ and
the vertical axis the corresponding angle. Starting from the common baseline graph described above,
we run a statistic-driven degree-preserving two-switch rewiring scheme designed to induce
disassortativity in a chosen endpoint-degree correlation.

For a directed edge $u\to v$ and a chosen focus, define the edge-endpoint data pair
$(X_{u\to v},Y_{u\to v})$ by
\[
(X_{u\to v},Y_{u\to v})
\in
\Bigl\{
\bigl(d_u^{\mathrm{in}},d_v^{\mathrm{in}}\bigr),\;
\bigl(d_u^{\mathrm{in}},d_v^{\mathrm{out}}\bigr),\;
\bigl(d_u^{\mathrm{out}},d_v^{\mathrm{in}}\bigr),\;
\bigl(d_u^{\mathrm{out}},d_v^{\mathrm{out}}\bigr)
\Bigr\}
\]
The corresponding assortativity coefficient is the Pearson correlation across edges,
\[
r(\text{focus})
:=
\mathrm{Corr}\Bigl(X_{u\to v},Y_{u\to v}:\;(u\to v)\in E\Bigr)
\]
Rewiring is performed in passes that repeatedly select pairs of edges with strongly contrasting
endpoint-degree scores and apply feasible two-switches that most strongly decrease the targeted
correlation. Edges are ranked by the focus-dependent degree product $X_{u\to v}Y_{u\to v}$, and
high-score edges are paired with low-score edges. A candidate two-switch is accepted only if it
decreases the centered-degree numerator
\[
\sum_{(u\to v)\in E}(X_{u\to v}-\mu_X)(Y_{u\to v}-\mu_Y)
\]
by at least a threshold $\tau$ (default $\tau=0$). Participation caps are enforced throughout by
restricting how often any vertex can be used in accepted swaps, implemented through vertex-disjoint
batches within each pass, in line with the bounded-participation logic used in the paper.

\paragraph{Interpreting the four assortativity patterns \\}

The neutral baseline suggests $\mathbf v^{\mathrm R}\approx \mathbf d^{\mathrm{out}}$ and
$\mathbf v^{\mathrm L}\approx \mathbf d^{\mathrm{in}}$; see
Proposition~\ref{prop:neutral-degree-evec} and the discussion around
Definition~\ref{def:angular-distance}. Assortativity rewiring breaks this neutrality selectively,
because accepted swaps are conditioned on which degree type is being matched, or anti-matched, at
the source and target ends. When the rewiring directly distorts the degree variables that enter the
relevant eigenvector summations, the corresponding angle drifts; when it acts only through
variables outside those summations, the induced perturbations are closer to mean-zero noise and the
angle mainly wanders.

\medskip
For in--in assortativity, with
$r_{\mathrm{in\!-\!in}}=\mathrm{Corr}(d_u^{\mathrm{in}},d_v^{\mathrm{in}})$,
driving the correlation negative breaks high-in/high-in and low-in/low-in matching and replaces it
with high-in/low-in pairings. This coherently distorts the incoming structure while preserving
degrees. Since $\mathbf v^{\mathrm L}$ is anchored by incoming reinforcement in the neutral regime,
it moves away from the fixed in-degree profile $\mathbf d^{\mathrm{in}}$, producing a visible
drift in $\theta_L(\mathrm{in})$. Out-degree does not enter the acceptance rule, so the outgoing
changes are effectively degree-agnostic and act approximately as mean-zero perturbations relative
to $\mathbf d^{\mathrm{out}}$. Accordingly, $\theta_R(\mathrm{out})$ fluctuates only modestly.

\medskip
For out--out assortativity, with
$r_{\mathrm{out\!-\!out}}=\mathrm{Corr}(d_u^{\mathrm{out}},d_v^{\mathrm{out}})$,
the same logic applies with the roles reversed. Driving the correlation negative breaks
high-out/high-out and low-out/low-out matching and replaces it with high-out/low-out pairings,
thereby reorganizing outgoing capacity while leaving degrees unchanged. Since
$\mathbf v^{\mathrm R}$ is the eigenvector most sensitive to outgoing structure in the neutral
heuristic, it moves away from $\mathbf d^{\mathrm{out}}$, yielding a clear drift in
$\theta_R(\mathrm{out})$, while $\theta_L(\mathrm{in})$ remains close to baseline.

\medskip
For in--out assortativity, with
$r_{\mathrm{in\!-\!out}}=\mathrm{Corr}(d_u^{\mathrm{in}},d_v^{\mathrm{out}})$,
the rewiring creates a crossed relationship relative to the neutral fixed-point equations: the
right eigenvector depends on the out-degrees of targets, while the left eigenvector depends on the
in-degrees of sources. Driving $r_{\mathrm{in\!-\!out}}$ negative therefore reorganizes both which
sources are active and which targets are selected according to the very variables that enter the
eigenvector summations. Because both sides of the summation logic are directly affected, both
Perron vectors move away from their degree anchors, producing the observed joint drift in
$\theta_R(\mathrm{out})$ and $\theta_L(\mathrm{in})$.

\medskip
For out--in assortativity, with
$r_{\mathrm{out\!-\!in}}=\mathrm{Corr}(d_u^{\mathrm{out}},d_v^{\mathrm{in}})$,
the rewiring instead constrains the source side by out-degree and the target side by in-degree,
that is, by variables lying outside the core summation terms of the neutral fixed-point equations.
Driving $r_{\mathrm{out\!-\!in}}$ negative forces high-out sources to connect to low-in targets,
but it does not systematically alter the quantities actually being summed in the eigenvector
relations. The resulting structural changes therefore act approximately as mean-zero perturbations
relative to the expected sums, allowing both Perron vectors to remain close to their degree
anchors. As a result, neither angle exhibits a pronounced drift.

\subsection{Local-structure rewiring}

\begin{figure}[H]
  \centering
\scalebox{1}{\includegraphics[width=0.75\textwidth]{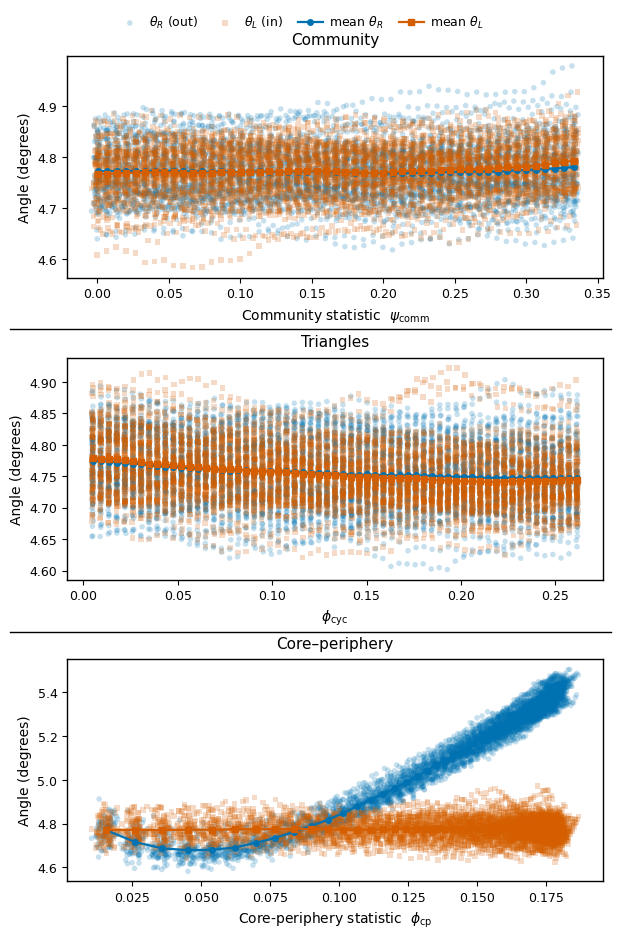}}
  \caption{Local-structure experiments. Top: communities ($\phi_{\mathrm{com}}$). Middle:
  triangles ($\phi_{3\text{-}\mathrm{cyc}}$). Bottom: core--periphery
  ($\phi_{\mathrm{cp}}$). Scatter plots show logged snapshots: solid curves show empirical mean
  angles across the Monte Carlo sample.}
  \label{fig:app-local}
\end{figure}

Figure~\ref{fig:app-local} plots the same two angles, $\theta_R(\mathrm{out})$ and
$\theta_L(\mathrm{in})$, against three statistics that reward local structure: community contrast,
triangle density, and core--periphery contrast. Dots are logged snapshots along rewiring
trajectories, and solid curves show empirical mean angles at each logging milestone across the
Monte Carlo sample. Starting again from the common baseline graph described above, we run a
degree-preserving two-switch Markov chain in which each accepted move is chosen to increase a
designated local-structure statistic. Angles are logged at regular milestones.

The three rewiring schemes are as follows. For communities
(Definition~\ref{def:community-stat}), vertices are randomly partitioned into two equally sized
groups, and cross-group edges are rewired into within-group edges so as to increase the community
contrast statistic $\phi_{\mathrm{com}}$. For triangles
(Definition~\ref{def:cycle_stat}), we target the directed $3$-cycle density
$\phi_{3\text{-}\mathrm{cyc}}$, accepting a local degree-preserving move only if it creates a
directed triangle without destroying an existing one. For core--periphery
(Definition~\ref{def:core-periphery-stat}), the core set $\mathcal H$ is defined as the top
$25\%$ of vertices by out-degree and held fixed during the run; rewiring selects
periphery--periphery edges and redirects them toward the core so as to reduce
$e_{\mathcal{LL}}$ and increase $\phi_{\mathrm{cp}}$.

These three experiments illustrate the wandering-versus-drift theme discussed in
Conjecture~\ref{conj:wandering_angle}: communities and triangles show little systematic drift,
whereas core--periphery rewiring does.

\paragraph{Communities \\}
The community two-switch is driven by group labels rather than by degrees. When a cross-group edge
is rewired into a within-group edge, the new target is chosen without regard to its in- or
out-degree. Consequently, the expected degrees of a node's targets and sources remain largely
unaltered. Since the rewiring does not systematically bias the variables that appear in the
eigenvector fixed-point summations, the induced perturbations act approximately as mean-zero noise
relative to the degree anchors. Both $\theta_R(\mathrm{out})$ and $\theta_L(\mathrm{in})$
therefore wander close to their neutral-baseline values, with only small graph-to-graph
heterogeneity.

\paragraph{Triangles \\}
Triangle-forming moves are likewise local and primarily topological. They enforce the existence of
directed $3$-cycles by checking for the presence or absence of specific edges, without conditioning
on the degrees of the nodes completing the cycle. Since the acceptance rule is degree-agnostic, it
does not persistently inflate or deflate the source- and target-degree terms appearing in the
fixed-point summations. Without a systematic bias in those sums, there is no persistent force
pushing Perron mass away from the degree vectors. This explains the near-flat mean curves and
contrasts sharply with the explicit drifts observed under targeted assortativity rewiring.

\paragraph{Core--periphery \\}
Core--periphery rewiring is different because it explicitly forces a relationship between topology
and degree size, with the core identified as the set of high out-degree vertices. By routing edges
away from $\mathcal L\to\mathcal L$ and toward the core, the rewiring systematically raises the
out-degrees of the targets seen by periphery nodes. In the right-eigenvector fixed-point equation,
this boosts the relevant summation term for periphery nodes relative to their own out-degrees,
breaking the proportionality between $\mathbf v^{\mathrm R}$ and $\mathbf d^{\mathrm{out}}$. This
targeted manipulation produces the visible drift in $\theta_R(\mathrm{out})$. By contrast,
$\theta_L(\mathrm{in})$ remains comparatively stable because the procedure does not symmetrically
condition on the in-degrees of sources, so the incoming-side summation is much less distorted.

\newpage
\singlespacing
\footnotesize
\bibliographystyle{plainnat}
\bibliography{eigen.bib}
\end{document}